\documentclass[10pt,conference]{IEEEtran}
\usepackage{tikz}
\usetikzlibrary{quantikz}
\usetikzlibrary{shadings}
\usepackage{braket}
\usepackage{amsmath, amsthm, amssymb}
\usepackage{bbm}
\theoremstyle{definition}

\newtheorem*{remark}{Remark}
\usepackage{cite}
\usepackage{hyperref}
\def\BibTeX{{\rm B\kern-.05em{\sc i\kern-.025em b}\kern-.08em

    T\kern-.1667em\lower.7ex\hbox{E}\kern-.125emX}}
\begin{document}

\title{Decomposition Algorithm of an Arbitrary Pauli Exponential through a Quantum Circuit }


\author{\IEEEauthorblockN{Maximilian Balthasar Mansky, Victor Ramos Puigvert, Santiago Londoño Castillo, Claudia Linnhoff-Popien}
\IEEEauthorblockA{LMU Munich\\
Munich, Germany\\
Email: maximilian-balthasar.mansky@ifi.lmu.de}}

\maketitle

\begin{abstract}
We review the staircase algorithm to decompose the exponential of a generalized Pauli matrix and we propose two alternative recursive methods which offer more efficient quantum circuits. The first algorithm we propose, defined as the inverted staircase algorithm, is more efficient in comparison to  the standard staircase algorithm in the number of one-qubit gates, giving a polynomial improvement of $n/2$. For our second algorithm, we introduce fermionic $\operatorname{SWAP}$ quantum gates and a systematic way of generalizing these. Such fermionic gates offer a simplification of the number of quantum gates, in particular of $\operatorname{CNOT}$ gates, in most quantum circuits. Regarding the staircase algorithm, fermionic quantum gates reduce the number of $\operatorname{CNOT}$ gates in roughly $n/2$ for a large number of qubits. In the end, we discuss the difference between the probability outcomes of fermionic and non-fermionic gates and show that, in general, due to interference, one cannot substitute fermionic gates through non-fermionic gates without altering the outcome of the circuit.
\end{abstract}

 \begin{IEEEkeywords}
Quantum Computing, Staircase Algorithm, Quantum Decomposition, Hamiltonian Dynamics, Fermionic Quantum Gates, Pauli Exponential.
\end{IEEEkeywords}

\section{Introduction}
\label{sec:introduction}

Quantum computers allow the application of the research developments of quantum mechanics towards a computational approach. In recent years there has been much progress in the implementation of the technology on physical hardware \cite{medicine_quantum_2019}, but issues of size and fidelity still mar the computers. Current quantum computers can only hold a small amount of information-carrying units (qubits) and address them with limited accuracy. This era of noisy small- and medium-scale hardware has been dubbed the NISQ era \cite{preskill_quantum_2018}. 

Quantum computing is expected to find applications across a range of different topics, including cryptography, many-body physics simulations, search and optimization, and game theory \cite{montanaro_quantum_2016}. Most likely additional application areas will be found, where the advantages of quantum computing can come into play.

Beyond physical limitations,  algorithms that can run efficiently on quantum hardware are also under constant development. So far, three main quantum algorithms have been discovered, namely, Shor \cite{shor_algorithms_1994}, Grover \cite{grover_fast_1996} and Deutsch-Josza \cite{deutsch_rapid_1992}. The search for new algorithms is proceeding slowly, but there is the belief that new algorithms can be found \cite{shor_progress_2004}. 

Moreover, there are a number of `works in practice' algorithms such as quantum machine learning \cite{schuld_quantum_2019} and variational approaches \cite{schuld_introduction_2015, biamonte_quantum_2017, benedetti_variational_2021}. The idea behind these approaches is to have a classical control loop in place that updates the parameters of the circuit and tunes it to produce a particular outcome, in direct analogy to machine learning. Starting from a fixed circuit structure (``Ansatz") with parametrized gates, the parameters are updated until a loss function is small enough \cite{schuld_introduction_2015}. 

These approaches hinge on building quantum circuits out of individual building blocks according to some scheme, in a forward construction. These building blocks are called rotation gates and $n$-qubit gates, depending if they act on an individual qubit or on several qubits at once, and both are implementable on quantum computers \cite{nielsen2002quantum}. Furthermore, by using these building blocks, it is possible to simulate through quantum circuits a general interaction of the type $e^{-iH t}$, where $H=h_i\sigma_I$ is an arbitrary Hamiltonian \cite{nielsen2002quantum}. This method for simulating a Hamiltonian is very similar to the so-called staircase algorithm, in which one adds two $\operatorname{CNOT}$ gates and a few one-qubit quantum gates for each additional Pauli matrix in the Hamiltonian surrounding a base rotation matrix $R_z$. 

In this paper, we review the staircase algorithm in detail in section \ref{sec:staircase_algorithm} and, furthermore, we propose two more efficient algorithms to simulate the interaction of arbitrary Hamiltonians of the type $H=h_i\sigma_I$, described in section \ref{sec:inversestaircase} and \ref{sec:algorithm}. The first algorithm is equivalent to the staircase algorithm, where we only change the base rotation matrix from $R_z$ to $R_x$, and the standard $\operatorname{CNOT}$ gates to \textit{inverted} ones. This algorithm, which we define as the \textit{inverted} staircase algorithm, is described in detail in section \ref{sec:inversestaircase}. It theoretically gives the same number of $\operatorname{CNOT}$ gates but improves the required number of one-qubit gates. 

In section \ref{sec:gates} we introduce fermionic $\operatorname{SWAP}$ quantum gates, which were defined first in \cite{cirac2009fswap}. For the second algorithm, we propose a method to systematically enlarge standard $\operatorname{SWAP}$ gates and $\operatorname{CNOT}$ gates through fermionic $\operatorname{SWAP}$ gates. Such enlarged fermionic quantum gates, if implementable, reduce the number of $\operatorname{CNOT}$ gates required to simulate the exponential of a generalized Pauli matrix through the inverted staircase algorithm, see section \ref{sec:algorithm}. There we discuss the expansion of the inverted staircase algorithm through fermionic gates. We also include some explicit examples in section \ref{sec:examples} to try to make clearer the implementation of enlarged fermionic gates.

Moreover, we also assess the performance of both of our algorithms with respect to the number of one-qubit gates and $\operatorname{CNOTs}$. The calculations comparing the efficiency of these three algorithms are performed in section \ref{sec:comparison}.

In section \ref{sec:trotter}, we discuss some potential applications of the proposed algorithms for quantum simulations. Since the terms composing a Hamiltonian do, in general, not commute, to simulate the dynamics of arbitrary Hamiltonians it is necessary to use a decomposition method, \cite{Osborne_2012}. Through the Suzuki-Trotter decomposition, we show how our proposed algorithms can be combined with this decomposition method for determining the quantum circuit decomposing the exponential of an arbitrary Hamiltonian up to a certain error. Furthermore, we review some relevant Hamiltonians which can be simulated through quantum circuits and have direct applications in physics. Finally, in the last section, we examine the difference between the probability outcomes of fermionic and non-fermionic quantum gates, see section \ref{sec:measurement}.

\section{Fermionic Gate System}
\label{sec:gates}
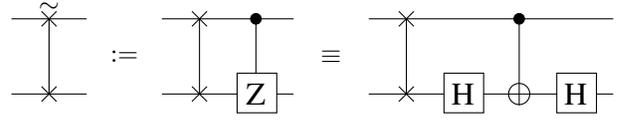
\begin{figure}[h!]
        \centering
        \begin{tikzpicture}
    \draw (6.5,8)--(7.5,8);
    \draw (6.5,7)--(7.5,7);

    \node at (7,8.15){$\sim$};
    \draw (6.9,8.1)--(7.1,7.9);
    \draw (6.9,7.9)--(7.1,8.1);
    \draw (6.9,7.1)--(7.1,6.9);
    \draw (6.9,6.9)--(7.1,7.1);
    \draw (7,8)--(7,7);

    \node at (8,7.5){$:=$};

    \draw (8.5,8)--(10.25,8);
    \draw (8.5,7)--(10.25,7);
    
    \draw (8.9,8.1)--(9.1,7.9);
    \draw (8.9,7.9)--(9.1,8.1);
    \draw (8.9,7.1)--(9.1,6.9);
    \draw (8.9,6.9)--(9.1,7.1);
    \draw (9,8)--(9,7);

    \draw (9.75,8)--(9.75,7);
    \filldraw (9.75,8) circle (2pt);
    \filldraw[draw=black, fill=white] (9.5,6.75) rectangle ++(15pt,15pt);
    \node at (9.75,7){\large Z};

    \node at (10.75,7.5){$\equiv$};

    \draw (11.25,8)--(14.5,8);
    \draw (11.25,7)--(14.5,7);

    \draw (11.65,8.1)--(11.85,7.9);
    \draw (11.65,7.9)--(11.85,8.1);
    \draw (11.65,7.1)--(11.85,6.9);
    \draw (11.65,6.9)--(11.85,7.1);
    \draw (11.75,8)--(11.75,7);

    \filldraw[draw=black, fill=white] (12.25,6.75) rectangle ++(15pt,15pt);
    \node at (12.5,7){\large H};

    \draw (13.25,7) circle (4pt);
    \draw (13.25,6.85)--(13.25,8);
    \filldraw (13.25,8) circle (2pt);

    \filldraw[draw=black, fill=white] (13.75,6.75) rectangle ++(15pt,15pt);
    \node at (14,7){\large H};
\end{tikzpicture}
	\caption{Definition of a fermionic $\operatorname{SWAP}$ which amounts to a $\operatorname{SWAP}$ gate followed or preceded by a controlled-Z gate. We display a fermionic gate through a tilde sign above the standard quantum gate.}
	\label{fig:FSWAP}
\end{figure}

In quantum computation, a standard $\operatorname{SWAP}$ gate \textit{swaps} the states of two qubits, $\ket{a,b}\to\ket{b,a}$, and is defined through three $\operatorname{CNOT}$ gates, \cite{nielsen2002quantum}. In particular, let $\ket{\psi}$ be a vector state on $SU(4)$ represented by the basis $\ket{00},\ket{01},\ket{10},\ket{11}$:
\begin{equation}
\ket{\psi}=\alpha_{00}\ket{00}+\alpha_{01}\ket{01}+\alpha_{10}\ket{10}+\alpha_{11}\ket{11}
\end{equation}
Then, the action of a $\operatorname{SWAP}$ gate on $\ket{\psi}$ is given by
\begin{equation}
\operatorname{SWAP}\ket{\psi}=\alpha_{00}\ket{00}+\alpha_{10}\ket{01}
        +\alpha_{01}\ket{10}+\alpha_{11}\ket{11}.
\end{equation}
where the placement of the prefactors $\alpha_{01}$ and $\alpha_{10}$ has been interchanged. This action can be represented with respect to the previously-introduced basis by the following matrix:
\begin{equation}
    \operatorname{SWAP}:=\begin{pmatrix}
    1&0&0&0\\
    0&0&1&0\\
    0&1&0&0\\
    0&0&0&1\end{pmatrix}.
\end{equation}
In \cite{cirac2009fswap}, a similar gate to the $\operatorname{SWAP}$ gate, the so-called fermionic $\operatorname{SWAP}$ gate, was introduced to account for the minus sign that arises when two fermionic modes are exchanged. The action of the fermionic $\operatorname{SWAP}$ gate on $\ket{\psi}$ was defined as
\begin{equation}
\operatorname{SWAP}\ket{\psi}=\alpha_{00}\ket{00}+\alpha_{10}\ket{01}
        +\alpha_{01}\ket{10}-\alpha_{11}\ket{11},
\end{equation}
which, in the same basis as before, can be represented by the following matrix:
\begin{equation}
    \operatorname{SWAP}:=\begin{pmatrix}
    1&0&0&0\\
    0&0&1&0\\
    0&1&0&0\\
    0&0&0&-1\end{pmatrix}.
\end{equation}
This fermionic $\operatorname{SWAP}$ gate can be expressed through the circuit corresponding to a $\operatorname{SWAP}$ gate followed or preceded by a controlled-Z gate, see figure \ref{fig:FSWAP}.

In quantum computation theory, gates can be enlarged systematically through $\operatorname{SWAP}$ gates. This allows the construction of non-adjacent quantum gates, such as a $\operatorname{CNOT}$ gate acting on the first and the third qubit. If implementable, such extended quantum gates imply a drastic reduction in the number of gates required to synthesize a desired circuit. Similarly, the successive generation of fermionic $\operatorname{SWAP}$ gates follows a recursive algorithm that allows the creation of enlarged fermionic $\operatorname{SWAP}$ gates and enlarged fermionic $\operatorname{CNOT}$ gates. The enlargement works as follows: if a $\operatorname{SWAP}$ or an $\operatorname{fSWAP}$ quantum gate gets enlarged through $\operatorname{SWAP}$ gates then the signs of its dominant off-diagonal entries get copied, while if it gets enlarged through $\operatorname{fSWAP}$ gates the signs of its dominant off-diagonal entries get reversed, see figures \ref{fig:fSWAP(+-)}, \ref{fig:fSWAP(++)}, \ref{fig:SWAP(++)}, and \ref{fig:SWAP(+-)}. We display the signs of the off-diagonal elements to keep track of the negative signs. 

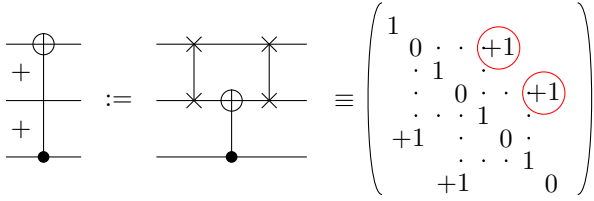
\begin{figure}[h!]
        \centering
        \begin{tikzpicture}

    \draw (4.5,8)--(5.5,8);
    \draw (4.5,7.25)--(5.5,7.25);
    \draw (4.5,6.5)--(5.5,6.5);

    \draw (5,8) circle (4pt);
    \filldraw (5,6.5) circle (2pt);
    \draw (5,8.15)--(5,6.5);
    \node at (4.7,7.63){$+$};
    \node at (4.7,6.87){$+$};

    \node at (6,7.25){$:=$};

    \draw (6.5,8)--(8.5,8);
    \draw (6.5,7.25)--(8.5,7.25);
    \draw (6.5,6.5)--(8.5,6.5);

    \draw (6.9,8.1)--(7.1,7.9);
    \draw (6.9,7.9)--(7.1,8.1);
    \draw (6.9,7.35)--(7.1,7.15);
    \draw (6.9,7.15)--(7.1,7.35);
    \draw (7,8)--(7,7.25);

    \draw (7.5,7.25) circle (4pt);
    \filldraw (7.5,6.5) circle (2pt);
    \draw (7.5,7.4)--(7.5,6.5);

    \draw (7.9,8.1)--(8.1,7.9);
    \draw (7.9,7.9)--(8.1,8.1);
    \draw (7.9,7.35)--(8.1,7.15);
    \draw (7.9,7.15)--(8.1,7.35);
    \draw (8,8)--(8,7.25);

    \node at (9,7.25){$\equiv$};

    \node at (9.65,8.25){$1$};
    \node at (9.95,7.95){$0$};
    \node at (10.25,7.65){$1$};
    \node at (10.55,7.35){$0$};
    \node at (10.85,7.05){$1$};
    \node at (11.15,6.75){$0$};
    \node at (11.45,6.45){$1$};
    \node at (11.75,6.15){$0$};

    \node at (11.05,7.95){$+1$};
    \node at (9.85,6.75){$+1$};
    \node at (11.65,7.35){$+1$};
    \node at (10.45,6.15){$+1$};

    \draw[red] (11.05,7.95) circle (8pt);
    \draw[red] (11.65,7.35) circle (8pt);
    
    \filldraw (10.55,7.95) circle (.3pt);
    \filldraw (10.25,7.95) circle (.3pt);
    \filldraw (10.85,7.95) circle (.3pt);

    \filldraw (9.95,7.65) circle (.3pt);
    \filldraw (9.95,7.35) circle (.3pt);
    \filldraw (9.95,7.05) circle (.3pt);

    \filldraw (10.25,7.05) circle (.3pt);
    \filldraw (10.55,7.05) circle (.3pt);

    \filldraw (10.85,7.35) circle (.3pt);

    \filldraw (11.15,7.35) circle (.3pt);

    \filldraw (11.45,7.05) circle (.3pt);
    \filldraw (11.45,6.75) circle (.3pt);
    \filldraw (11.45,7.35) circle (.3pt);

    \filldraw (11.15,6.45) circle (.3pt);
    \filldraw (10.85,6.45) circle (.3pt);
    \filldraw (10.55,6.45) circle (.3pt);

    \filldraw (10.55,6.75) circle (.3pt);

    \filldraw (10.85,7.65) circle (.3pt);

    \draw (9.5,8.55) to [out=180, in=180, looseness=.25] ($(9.5,8.5)!1!(9.5,6)$);
    \draw (12.1,8.55) to [out=0, in=0, looseness=.25] ($(12.1,8.5)!1!(12.1,6)$);

\end{tikzpicture}
	\caption{Definition of a $\operatorname{CNOT}(+,+)$ gate between the first and the third qubit, which amounts to the traditional enlarged and inverted $\operatorname{CNOT}(3,1)$.}
	\label{fig:CNOT(++)}
\end{figure}

\begin{figure}[h!]
        \centering
        \begin{tikzpicture}

    \draw (4.5,8)--(5.5,8);
    \draw (4.5,7.25)--(5.5,7.25);
    \draw (4.5,6.5)--(5.5,6.5);

    \draw (5,8) circle (4pt);
    \filldraw (5,6.5) circle (2pt);
    \draw (5,8.15)--(5,6.5);
    \node at (4.7,7.63){$+$};
    \node at (4.7,6.87){$-$};

    \node at (6,7.25){$:=$};

    \draw (6.5,8)--(8.5,8);
    \draw (6.5,7.25)--(8.5,7.25);
    \draw (6.5,6.5)--(8.5,6.5);

    \node at (7,8.2){$\sim$};
    \draw (6.9,8.1)--(7.1,7.9);
    \draw (6.9,7.9)--(7.1,8.1);
    \draw (6.9,7.35)--(7.1,7.15);
    \draw (6.9,7.15)--(7.1,7.35);
    \draw (7,8)--(7,7.25);

    \draw (7.5,7.25) circle (4pt);
    \filldraw (7.5,6.5) circle (2pt);
    \draw (7.5,7.4)--(7.5,6.5);

    \node at (8,8.2){$\sim$};
    \draw (7.9,8.1)--(8.1,7.9);
    \draw (7.9,7.9)--(8.1,8.1);
    \draw (7.9,7.35)--(8.1,7.15);
    \draw (7.9,7.15)--(8.1,7.35);
    \draw (8,8)--(8,7.25);

    \node at (9,7.25){$\equiv$};

    \node at (9.65,8.25){$1$};
    \node at (9.95,7.95){$0$};
    \node at (10.25,7.65){$1$};
    \node at (10.55,7.35){$0$};
    \node at (10.85,7.05){$1$};
    \node at (11.15,6.75){$0$};
    \node at (11.45,6.45){$1$};
    \node at (11.75,6.15){$0$};

    \node at (11.05,7.95){$+1$};
    \node at (9.85,6.75){$+1$};
    \node at (11.65,7.35){$-1$};
    \node at (10.45,6.15){$-1$};

\draw[red] (11.05,7.95) circle (8pt);
\draw[blue] (11.65,7.35) circle (8pt);

    \filldraw (10.55,7.95) circle (.3pt);
    \filldraw (10.25,7.95) circle (.3pt);
    \filldraw (10.85,7.95) circle (.3pt);

    \filldraw (9.95,7.65) circle (.3pt);
    \filldraw (9.95,7.35) circle (.3pt);
    \filldraw (9.95,7.05) circle (.3pt);

    \filldraw (10.25,7.05) circle (.3pt);
    \filldraw (10.55,7.05) circle (.3pt);

    \filldraw (10.85,7.35) circle (.3pt);

    \filldraw (11.15,7.35) circle (.3pt);

    \filldraw (11.45,7.05) circle (.3pt);
    \filldraw (11.45,6.75) circle (.3pt);
    \filldraw (11.45,7.35) circle (.3pt);

    \filldraw (11.15,6.45) circle (.3pt);
    \filldraw (10.85,6.45) circle (.3pt);
    \filldraw (10.55,6.45) circle (.3pt);

    \filldraw (10.55,6.75) circle (.3pt);

    \filldraw (10.85,7.65) circle (.3pt);

    \draw (9.5,8.55) to [out=180, in=180, looseness=.25] ($(9.5,8.5)!1!(9.5,6)$);
    \draw (12.1,8.55) to [out=0, in=0, looseness=.25] ($(12.1,8.5)!1!(12.1,6)$);
    
\end{tikzpicture}
	\caption{Definition of a $\operatorname{CNOT}(+,-)$ gate between the first and the third qubit. Through enlarging with fermionic $\operatorname{SWAP}$ gates the signs of the dominant diagonal entry get reversed so that a negative sign appears in contrast to the standard $\operatorname{CNOT}(3,1)$.}
	\label{fig:CNOT(+-)}
\end{figure}
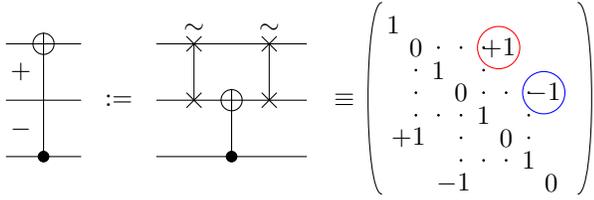

Note that, the difference between these enlarged quantum gates lies in some sign-prefactor of some off-diagonal elements and not in the exchange of the qubits itself. The enlargement of standard $\operatorname{CNOT}$ gates follows the same recursion: if it gets enlarged through $\operatorname{SWAP}$ gates then the signs of its dominant off-diagonal entries get copied, while if it gets enlarged through $\operatorname{fSWAP}$ gates they get reversed, see figures \ref{fig:CNOT(++)} and \ref{fig:CNOT(+-)}.

\begin{figure}[h!]
        \centering
        \begin{tikzpicture}

    \draw (4.5,8)--(5.5,8);
    \draw (4.5,7.25)--(5.5,7.25);
    \draw (4.5,6.5)--(5.5,6.5);

    \node at (5,8.15){$\sim$};
    \draw (4.9,7.9)--(5.1,8.1);
    \draw (4.9,8.1)--(5.1,7.9);
    \draw (4.9,6.4)--(5.1,6.6);
    \draw (4.9,6.6)--(5.1,6.4);
    \draw (5,8)--(5,6.5);
    \node at (4.7,7.63){$+$};
    \node at (4.7,6.87){$-$};

    \node at (6,7.25){$:=$};

    \draw (6.5,8)--(8.5,8);
    \draw (6.5,7.25)--(8.5,7.25);
    \draw (6.5,6.5)--(8.5,6.5);

    \node at (7,8.2){$\sim$};
    \draw (6.9,8.1)--(7.1,7.9);
    \draw (6.9,7.9)--(7.1,8.1);
    \draw (6.9,7.35)--(7.1,7.15);
    \draw (6.9,7.15)--(7.1,7.35);
    \draw (7,8)--(7,7.25);

    \node at (7.5,7.4){$\sim$};
    \draw (7.4,7.15)--(7.6,7.35);
    \draw (7.4,7.35)--(7.6,7.15);
    \draw (7.4,6.4)--(7.6,6.6);
    \draw (7.4,6.6)--(7.6,6.4);
    \draw (7.5,7.25)--(7.5,6.5);

    \node at (8,8.2){$\sim$};
    \draw (7.9,8.1)--(8.1,7.9);
    \draw (7.9,7.9)--(8.1,8.1);
    \draw (7.9,7.35)--(8.1,7.15);
    \draw (7.9,7.15)--(8.1,7.35);
    \draw (8,8)--(8,7.25);

    \node at (9,7.25){$\equiv$};

    \node at (9.65,8.25){$1$};
    \node at (9.95,7.95){$0$};
    \node at (10.25,7.65){$1$};
    \node at (10.55,7.35){$0$};
    \node at (10.85,7.05){$0$};
    \node at (11.15,6.75){$-1$};
    \node at (11.45,6.45){$0$};
    \node at (11.75,6.15){$-1$};

    \draw (11.2,6.75) circle (8pt);
    \draw (11.8,6.15) circle (8pt);

    \node at (10.75,7.95){$+1$};
    \node at (9.85,7.05){$+1$};
    \node at (11.35,7.35){$-1$};
    \node at (10.45,6.45){$-1$};

    \draw[red] (10.75,7.95) circle (8pt);
    \draw[blue] (11.35,7.35) circle (8pt);

    \filldraw (10.55,7.95) circle (.3pt);
    \filldraw (10.25,7.95) circle (.3pt);
    
    \filldraw (9.95,7.65) circle (.3pt);
    \filldraw (9.95,7.35) circle (.3pt);

    \filldraw (10.25,7.05) circle (.3pt);
    \filldraw (10.55,7.05) circle (.3pt);

    \filldraw (10.85,7.35) circle (.3pt);

    \filldraw (11.15,7.35) circle (.3pt);

    \filldraw (11.45,7.05) circle (.3pt);
    \filldraw (11.45,6.75) circle (.3pt);

    \filldraw (11.15,6.45) circle (.3pt);
    \filldraw (10.85,6.45) circle (.3pt);

    \filldraw (10.55,6.75) circle (.3pt);

    \filldraw (10.85,7.65) circle (.3pt);

    \draw (9.5,8.55) to [out=180, in=180, looseness=.25] ($(9.5,8.5)!1!(9.5,6)$);
    \draw (12.1,8.55) to [out=0, in=0, looseness=.25] ($(12.1,8.5)!1!(12.1,6)$);

\end{tikzpicture}
	\caption{Definition of a $\operatorname{fSWAP}(+,-)$ gate between the first and the third qubit. The enlargement through $\operatorname{fSWAP}$ gates reverses the sign of its dominant off-diagonal entries. The tilde sign denotes that the base gate is an $\operatorname{fSWAP}$, which implies a negative sign on the lower-half diagonal entries.}
	\label{fig:fSWAP(+-)}
\end{figure}
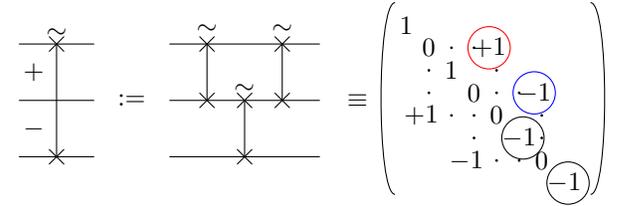
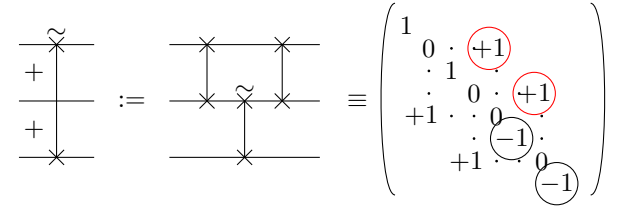
\begin{figure}[h!]
        \centering
	\begin{tikzpicture}

    \draw (4.5,8)--(5.5,8);
    \draw (4.5,7.25)--(5.5,7.25);
    \draw (4.5,6.5)--(5.5,6.5);

    \node at (5,8.15){$\sim$};
    \draw (4.9,7.9)--(5.1,8.1);
    \draw (4.9,8.1)--(5.1,7.9);
    \draw (4.9,6.4)--(5.1,6.6);
    \draw (4.9,6.6)--(5.1,6.4);
    \draw (5,8)--(5,6.5);
    \node at (4.7,7.63){$+$};
    \node at (4.7,6.87){$+$};

    \node at (6,7.25){$:=$};

    \draw (6.5,8)--(8.5,8);
    \draw (6.5,7.25)--(8.5,7.25);
    \draw (6.5,6.5)--(8.5,6.5);

    \draw (6.9,8.1)--(7.1,7.9);
    \draw (6.9,7.9)--(7.1,8.1);
    \draw (6.9,7.35)--(7.1,7.15);
    \draw (6.9,7.15)--(7.1,7.35);
    \draw (7,8)--(7,7.25);

    \node at (7.5,7.4){$\sim$};
    \draw (7.4,7.15)--(7.6,7.35);
    \draw (7.4,7.35)--(7.6,7.15);
    \draw (7.4,6.4)--(7.6,6.6);
    \draw (7.4,6.6)--(7.6,6.4);
    \draw (7.5,7.25)--(7.5,6.5);

    \draw (7.9,8.1)--(8.1,7.9);
    \draw (7.9,7.9)--(8.1,8.1);
    \draw (7.9,7.35)--(8.1,7.15);
    \draw (7.9,7.15)--(8.1,7.35);
    \draw (8,8)--(8,7.25);

    \node at (9,7.25){$\equiv$};

    \node at (9.65,8.25){$1$};
    \node at (9.95,7.95){$0$};
    \node at (10.25,7.65){$1$};
    \node at (10.55,7.35){$0$};
    \node at (10.85,7.05){$0$};
    \node at (11.05,6.75){$-1$};
    \node at (11.45,6.45){$0$};
    \node at (11.65,6.15){$-1$};

    \draw (11.05,6.75) circle (8pt);
    \draw (11.65,6.15) circle (8pt);

    \node at (10.75,7.95){$+1$};
    \node at (9.85,7.05){$+1$};
    \node at (11.35,7.35){$+1$};
    \node at (10.45,6.45){$+1$};

    \draw[red] (10.75,7.95) circle (8pt);
    \draw[red] (11.35,7.35) circle (8pt);

    \filldraw (10.55,7.95) circle (.3pt);
    \filldraw (10.25,7.95) circle (.3pt);
    
    \filldraw (9.95,7.65) circle (.3pt);
    \filldraw (9.95,7.35) circle (.3pt);

    \filldraw (10.25,7.05) circle (.3pt);
    \filldraw (10.55,7.05) circle (.3pt);

    \filldraw (10.85,7.35) circle (.3pt);

    \filldraw (11.15,7.35) circle (.3pt);

    \filldraw (11.45,7.05) circle (.3pt);
    \filldraw (11.45,6.75) circle (.3pt);

    \filldraw (11.15,6.45) circle (.3pt);
    \filldraw (10.85,6.45) circle (.3pt);

    \filldraw (10.55,6.75) circle (.3pt);

    \filldraw (10.85,7.65) circle (.3pt);

    \draw (9.5,8.55) to [out=180, in=180, looseness=.25] ($(9.5,8.5)!1!(9.5,6)$);
    \draw (12.1,8.55) to [out=0, in=0, looseness=.25] ($(12.1,8.5)!1!(12.1,6)$);
    
\end{tikzpicture}	
	\caption{Definition of a $\operatorname{fSWAP}(+,+)$ gate between the first and the third qubit. The enlargement through $\operatorname{SWAP}$ gates copies the symbol of its dominant off-diagonal entries. The tilde sign denotes that the base gate is an $\operatorname{fSWAP}$, which implies a negative sign on the lower-half diagonal entries.}
	\label{fig:fSWAP(++)}
\end{figure}
\begin{figure}[h!]
        \centering
	\begin{tikzpicture}

    \draw (4.5,8)--(5.5,8);
    \draw (4.5,7.25)--(5.5,7.25);
    \draw (4.5,6.5)--(5.5,6.5);

    \draw (4.9,7.9)--(5.1,8.1);
    \draw (4.9,8.1)--(5.1,7.9);
    \draw (4.9,6.4)--(5.1,6.6);
    \draw (4.9,6.6)--(5.1,6.4);
    \draw (5,8)--(5,6.5);
    \node at (4.7,7.63){$+$};
    \node at (4.7,6.87){$+$};

    \node at (6,7.25){$:=$};

    \draw (6.5,8)--(8.5,8);
    \draw (6.5,7.25)--(8.5,7.25);
    \draw (6.5,6.5)--(8.5,6.5);

    \draw (6.9,8.1)--(7.1,7.9);
    \draw (6.9,7.9)--(7.1,8.1);
    \draw (6.9,7.35)--(7.1,7.15);
    \draw (6.9,7.15)--(7.1,7.35);
    \draw (7,8)--(7,7.25);

    \draw (7.4,7.15)--(7.6,7.35);
    \draw (7.4,7.35)--(7.6,7.15);
    \draw (7.4,6.4)--(7.6,6.6);
    \draw (7.4,6.6)--(7.6,6.4);
    \draw (7.5,7.25)--(7.5,6.5);

    \draw (7.9,8.1)--(8.1,7.9);
    \draw (7.9,7.9)--(8.1,8.1);
    \draw (7.9,7.35)--(8.1,7.15);
    \draw (7.9,7.15)--(8.1,7.35);
    \draw (8,8)--(8,7.25);

    \node at (9,7.25){$\equiv$};

    \node at (9.65,8.25){$1$};
    \node at (9.95,7.95){$0$};
    \node at (10.25,7.65){$1$};
    \node at (10.55,7.35){$0$};
    \node at (10.85,7.05){$0$};
    \node at (11.15,6.75){$1$};
    \node at (11.45,6.45){$0$};
    \node at (11.75,6.15){$1$};

    \node at (10.75,7.95){$+1$};
    \node at (9.85,7.05){$+1$};
    \node at (11.35,7.35){$+1$};
    \node at (10.45,6.45){$+1$};

    \draw[red] (10.75,7.95) circle (8pt);
    \draw[red] (11.35,7.35) circle (8pt);

    \filldraw (10.55,7.95) circle (.3pt);
    \filldraw (10.25,7.95) circle (.3pt);
    
    \filldraw (9.95,7.65) circle (.3pt);
    \filldraw (9.95,7.35) circle (.3pt);

    \filldraw (10.25,7.05) circle (.3pt);
    \filldraw (10.55,7.05) circle (.3pt);

    \filldraw (10.85,7.35) circle (.3pt);

    \filldraw (11.15,7.35) circle (.3pt);

    \filldraw (11.45,7.05) circle (.3pt);
    \filldraw (11.45,6.75) circle (.3pt);

    \filldraw (11.15,6.45) circle (.3pt);
    \filldraw (10.85,6.45) circle (.3pt);

    \filldraw (10.55,6.75) circle (.3pt);

    \filldraw (10.85,7.65) circle (.3pt);

    \draw (9.5,8.55) to [out=180, in=180, looseness=.25] ($(9.5,8.5)!1!(9.5,6)$);
    \draw (12.1,8.55) to [out=0, in=0, looseness=.25] ($(12.1,8.5)!1!(12.1,6)$);
    
\end{tikzpicture}	
	\caption{Definition of a $\operatorname{SWAP}(+,+)$ gate between the first and the third qubit, which amounts to the traditional enlarged $\operatorname{SWAP}$(1,3). The enlargement through $\operatorname{SWAP}$ gates copies the symbol of its dominant off-diagonal entries.}
	\label{fig:SWAP(++)}
\end{figure}
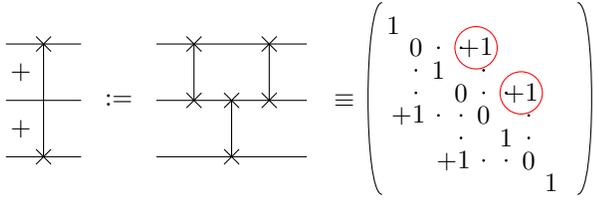
\begin{figure}[h!]
        \centering
	\begin{tikzpicture}

    \draw (4.5,8)--(5.5,8);
    \draw (4.5,7.25)--(5.5,7.25);
    \draw (4.5,6.5)--(5.5,6.5);

    \draw (4.9,7.9)--(5.1,8.1);
    \draw (4.9,8.1)--(5.1,7.9);
    \draw (4.9,6.4)--(5.1,6.6);
    \draw (4.9,6.6)--(5.1,6.4);
    \draw (5,8)--(5,6.5);
    \node at (4.7,7.63){$+$};
    \node at (4.7,6.87){$-$};

    \node at (6,7.25){$:=$};

    \draw (6.5,8)--(8.5,8);
    \draw (6.5,7.25)--(8.5,7.25);
    \draw (6.5,6.5)--(8.5,6.5);

    \node at (7,8.2){$\sim$};
    \draw (6.9,8.1)--(7.1,7.9);
    \draw (6.9,7.9)--(7.1,8.1);
    \draw (6.9,7.35)--(7.1,7.15);
    \draw (6.9,7.15)--(7.1,7.35);
    \draw (7,8)--(7,7.25);

    \draw (7.4,7.15)--(7.6,7.35);
    \draw (7.4,7.35)--(7.6,7.15);
    \draw (7.4,6.4)--(7.6,6.6);
    \draw (7.4,6.6)--(7.6,6.4);
    \draw (7.5,7.25)--(7.5,6.5);

    \node at (8,8.2){$\sim$};
    \draw (7.9,8.1)--(8.1,7.9);
    \draw (7.9,7.9)--(8.1,8.1);
    \draw (7.9,7.35)--(8.1,7.15);
    \draw (7.9,7.15)--(8.1,7.35);
    \draw (8,8)--(8,7.25);

    \node at (9,7.25){$\equiv$};

    \node at (9.65,8.25){$1$};
    \node at (9.95,7.95){$0$};
    \node at (10.25,7.65){$1$};
    \node at (10.55,7.35){$0$};
    \node at (10.85,7.05){$0$};
    \node at (11.15,6.75){$1$};
    \node at (11.45,6.45){$0$};
    \node at (11.75,6.15){$1$};

    \node at (10.75,7.95){$+1$};
    \node at (9.85,7.05){$+1$};
    \node at (11.35,7.35){$-1$};
    \node at (10.45,6.45){$-1$};

    \draw[red] (10.75,7.95) circle (8pt);
    \draw[blue] (11.35,7.35) circle (8pt);

    \filldraw (10.55,7.95) circle (.3pt);
    \filldraw (10.25,7.95) circle (.3pt);
    
    \filldraw (9.95,7.65) circle (.3pt);
    \filldraw (9.95,7.35) circle (.3pt);

    \filldraw (10.25,7.05) circle (.3pt);
    \filldraw (10.55,7.05) circle (.3pt);

    \filldraw (10.85,7.35) circle (.3pt);

    \filldraw (11.15,7.35) circle (.3pt);

    \filldraw (11.45,7.05) circle (.3pt);
    \filldraw (11.45,6.75) circle (.3pt);

    \filldraw (11.15,6.45) circle (.3pt);
    \filldraw (10.85,6.45) circle (.3pt);

    \filldraw (10.55,6.75) circle (.3pt);

    \filldraw (10.85,7.65) circle (.3pt);

    \draw (9.5,8.55) to [out=180, in=180, looseness=.25] ($(9.5,8.5)!1!(9.5,6)$);
    \draw (12.1,8.55) to [out=0, in=0, looseness=.25] ($(12.1,8.5)!1!(12.1,6)$);
\end{tikzpicture}	
	\caption{Definition of a $\operatorname{SWAP}(+,-)$ gate between the first and the third qubit. The enlargement through $\operatorname{fSWAP}$ gates reverses the sign of its dominant off-diagonal entries.}
	\label{fig:SWAP(+-)}
\end{figure}
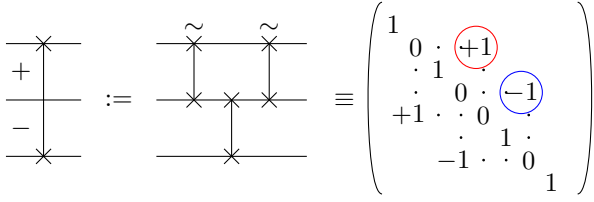

\section{Staircase Algorithm}
\label{sec:staircase_algorithm}
The basic algorithm most commonly used to determine the quantum circuit decomposing an exponential of a generalized Pauli matrix is the so-called staircase algorithm. This algorithm can be used, for instance, to simulate the dynamics of a Hamiltonian, see \cite{nielsen2002quantum}, \cite{whitfield2011simulation}. 

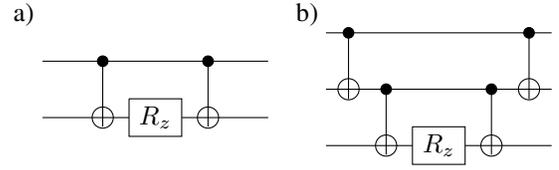
\begin{figure}[h!]
        \centering
        \begin{tikzpicture}

            \filldraw[draw=white, fill=white] (5.15,6.37)--(5.85,6.37)--(5.85,5.87)--(5.15,5.87)--(5.15,6.37);
            \node at (3.75,7.87){a)};
		\draw (4,7.25)--(7,7.25);
            \draw (4,6.5)--(7,6.5);

            \draw (4.8,6.5) circle (4pt);
            \draw (6.2,6.5) circle (4pt);
            \filldraw (4.8,7.25) circle (2pt);
		\filldraw (6.2,7.25) circle (2pt);
            \draw (4.8,7.25)--(4.8,6.38);
		\draw (6.2,7.25)--(6.2,6.38);
		\filldraw[draw=black, fill=white] (5.15,6.75)--(5.85,6.75)--(5.85,6.25)--(5.15,6.25)--(5.15,6.75);
            \node at (5.5,6.5){$R_z$};

	\end{tikzpicture}
        \begin{tikzpicture}
            \node at (3.75,8.25){b)};
		\draw (4,8)--(7,8);
		\draw (4,7.25)--(7,7.25);
            \draw (4,6.5)--(7,6.5);

		\draw (4.3,7.25) circle (4pt);
		\draw (6.7,7.25) circle (4pt);
		\filldraw (4.3,8) circle (2pt);
		\filldraw (6.7,8) circle (2pt);
		\draw (4.3,8)--(4.3,7.12);
		\draw (6.7,8)--(6.7,7.12);
            \draw (4.8,6.5) circle (4pt);
            \draw (6.2,6.5) circle (4pt);
            \filldraw (4.8,7.25) circle (2pt);
		\filldraw (6.2,7.25) circle (2pt);
            \draw (4.8,7.25)--(4.8,6.38);
		\draw (6.2,7.25)--(6.2,6.38);
		\filldraw[draw=black, fill=white] (5.15,6.75)--(5.85,6.75)--(5.85,6.25)--(5.15,6.25)--(5.15,6.75);
            \node at (5.5,6.5){$R_z$};

	\end{tikzpicture}
	\caption{Quantum circuits depicting how the staircase algorithm works. \textbf{a)} Quantum circuit decomposing the exponential $e^{-ia\sigma_z^1\otimes\sigma_z^2}$. \textbf{b)} Quantum circuit decomposing the exponential $e^{-ia\sigma_z^1\otimes\sigma_z^2\otimes\sigma_z^3}$. The addition of another $\otimes\sigma_z^i$ in the generalized Pauli matrix amounts to extending the quantum circuit with one $\operatorname{CNOT}$ gate at each side.}
	\label{fig:staircase}
\end{figure}

The staircase algorithm starts decomposing the exponential of the Hamiltonian $H=a\sigma_z^1\otimes\sigma_z^2$ through two $\operatorname{CNOT}$ gates surrounding a one-qubit $R_z(2a)$ gate. Then, for every new tensor product with a $\sigma_z^i$ it adds a pair of $\operatorname{CNOT}$ gates, one at each side of the one-qubit gate, see figure \ref{fig:staircase}. 

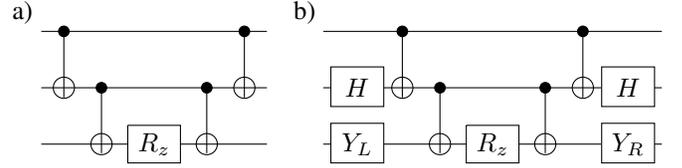
\begin{figure}[h!]
        \centering
        \begin{tikzpicture}
            \node at (3.75,8.25){a)};
		\draw (4,8)--(7,8);
		\draw (4,7.25)--(7,7.25);
            \draw (4,6.5)--(7,6.5);

		\draw (4.3,7.25) circle (4pt);
		\draw (6.7,7.25) circle (4pt);
		\filldraw (4.3,8) circle (2pt);
		\filldraw (6.7,8) circle (2pt);
		\draw (4.3,8)--(4.3,7.12);
		\draw (6.7,8)--(6.7,7.12);
            \draw (4.8,6.5) circle (4pt);
            \draw (6.2,6.5) circle (4pt);
            \filldraw (4.8,7.25) circle (2pt);
		\filldraw (6.2,7.25) circle (2pt);
            \draw (4.8,7.25)--(4.8,6.38);
		\draw (6.2,7.25)--(6.2,6.38);
		\filldraw[draw=black, fill=white] (5.15,6.75)--(5.85,6.75)--(5.85,6.25)--(5.15,6.25)--(5.15,6.75);
            \node at (5.5,6.5){$R_z$};

            \node at (7.5,8.25){b)};
            \draw (7.75,8)--(12.25,8);
            \draw (7.75,7.25)--(12.25,7.25);
            \draw (7.75,6.5)--(12.25,6.5);
		\filldraw[draw=black, fill=white] (9.65,6.75)--(10.35,6.75)--(10.35,6.25)--(9.65,6.25)--(9.65,6.75);
            \node at (10,6.5){$R_z$};
            \draw (9.3,6.5) circle (4pt);
            \draw (10.7,6.5) circle (4pt);
            \filldraw (9.3,7.25) circle (2pt);
		\filldraw (10.7,7.25) circle (2pt);
            \draw (9.3,7.25)--(9.3,6.38);
		\draw (10.7,7.25)--(10.7,6.38);
            \draw (8.8,7.25) circle (4pt);
		\draw (11.2,7.25) circle (4pt);
		\filldraw (8.8,8) circle (2pt);
		\filldraw (11.2,8) circle (2pt);
		\draw (8.8,8)--(8.8,7.12);
		\draw (11.2,8)--(11.2,7.12);

            \filldraw[draw=black, fill=white] (7.85,7) rectangle ++(20pt,15pt);
            \node at (8.2,7.25){$H$};
            \filldraw[draw=black, fill=white] (7.85,6.25) rectangle ++(20pt,15pt);
            \node at (8.2,6.5){$Y_L$};

            \filldraw[draw=black, fill=white] (11.45,7) rectangle ++(20pt,15pt);
            \node at (11.8,7.25){$H$};
            \filldraw[draw=black, fill=white] (11.45,6.25) rectangle ++(20pt,15pt);
            \node at (11.8,6.5){$Y_R$};

	\end{tikzpicture}
	\caption{Quantum circuits decomposing two exponentials $e^{-iaH}$ depicting how the staircase algorithm works for all Pauli matrices. \textbf{a)} $H=\sigma_z^1\otimes\sigma_z^2\otimes\sigma_z^3$ \textbf{b)} $H=\sigma_z^1\otimes\sigma_x^2\otimes\sigma_y^3$, for which it uses the relations $H\sigma_x^iH=\sigma_z^i$ on the second qubit and $Y_L\sigma_y^iY_R=\sigma_z^i$ on the third one, where $Y_L=R_z\left(\frac{\pi}{2}\right)H$ and $Y_R=HR_z\left(-\frac{\pi}{2}\right)$.}
	\label{fig:ZZZvsZXY}
\end{figure}

Through the relations between the Pauli matrices
\begin{equation}
    H\sigma_xH=\sigma_z
    \label{eq:relation1}
\end{equation}

and

\begin{equation}
     R_z\left(\frac{\pi}{2}\right)\sigma_xR_z\left(-\frac{\pi}{2}\right)=\sigma_y
     \label{eq:relation2}
\end{equation}

where $H$ denotes the Hadamard gate and $R_z$ is a one-qubit rotation operator about the $z$-axis, it is possible to account for the tensoring of Pauli matrices different from $\sigma_z$. For every $\sigma_x^i$ it is tensored with, it adds a pair of $\operatorname{CNOT}$ gates as well as a pair of $H$ gates, acting on the external side of the $i$th qubit. Similarly, for every $\sigma_y^i$ it is tensored with, it adds a pair of $\operatorname{CNOT}$ gates as well as a pair of $Y_{L/R}$ gates, where each $Y_L=HR_z\left(-\frac{\pi}{2}\right)$ acts on the external left side of the $i$th qubit and each $Y_R=HR_z\left(\frac{\pi}{2}\right)$ acts on the external right side of the $i$th qubit, see figure \ref{fig:ZZZvsZXY}. 

For every $\mathbbm{1}_{2x2}^i$ matrix it is tensored with, it adds a pair of $\operatorname{SWAP}$ gates surrounding the one-qubit $R_z$ gate. This is equivalent to adding an unaffected $i$th-qubit and thus allows simulating Hamiltonians with non-adjacent Pauli matrices.

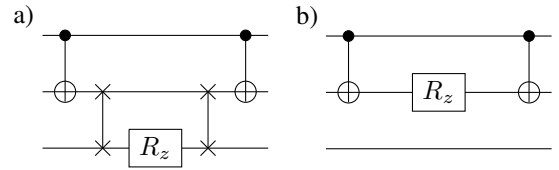
\begin{figure}[h!]
        \centering
        \begin{tikzpicture}
            \node at (3.75,8.25){a)};
		\draw (4,8)--(7,8);
		\draw (4,7.25)--(7,7.25);
            \draw (4,6.5)--(7,6.5);

		\draw (4.3,7.25) circle (4pt);
		\draw (6.7,7.25) circle (4pt);
		\filldraw (4.3,8) circle (2pt);
		\filldraw (6.7,8) circle (2pt);
		\draw (4.3,8)--(4.3,7.12);
		\draw (6.7,8)--(6.7,7.12);
            
            \draw (4.8,7.25)--(4.8,6.5);
		\draw (6.2,7.25)--(6.2,6.5);

            \draw (4.7,7.15)--(4.9,7.35);
            \draw (4.7,7.35)--(4.9,7.15);
            \draw (4.7,6.4)--(4.9,6.6);
            \draw (4.7,6.6)--(4.9,6.4);

            \draw (6.1,7.15)--(6.3,7.35);
            \draw (6.1,7.35)--(6.3,7.15);
            \draw (6.1,6.4)--(6.3,6.6);
            \draw (6.1,6.6)--(6.3,6.4);
		\filldraw[draw=black, fill=white] (5.15,6.75)--(5.85,6.75)--(5.85,6.25)--(5.15,6.25)--(5.15,6.75);
            \node at (5.5,6.5){$R_z$};

	\end{tikzpicture}
        \begin{tikzpicture}
            \filldraw[draw=white, fill=white] (5.15,6.75)--(5.85,6.75)--(5.85,6.25)--(5.15,6.25)--(5.15,6.75);
            
            \node at (3.75,8.25){b)};
		\draw (4,8)--(7,8);
		\draw (4,7.25)--(7,7.25);
            \draw (4,6.5)--(7,6.5);

		\draw (4.3,7.25) circle (4pt);
		\draw (6.7,7.25) circle (4pt);
		\filldraw (4.3,8) circle (2pt);
		\filldraw (6.7,8) circle (2pt);
		\draw (4.3,8)--(4.3,7.12);
		\draw (6.7,8)--(6.7,7.12);

		\filldraw[draw=black, fill=white] (5.15,7.5)--(5.85,7.5)--(5.85,7)--(5.15,7)--(5.15,7.5);
            \node at (5.5,7.25){$R_z$};

	\end{tikzpicture}
	\caption{For every $\mathbbm{1}_{2x2}^i$ matrix it is tensored with, it adds a pair of $\operatorname{SWAP}$ gates surrounding the one-qubit $R_z$ gate, which amounts to just adding an unaffected qubit. Both $\textbf{a)}$ and $\textbf{b)}$ quantum circuits simulate the same exponential $e^{-iaH}$, where $H=\sigma_z^1\otimes\sigma_z^2\otimes\mathbbm{1}^3$.}
	\label{fig:ZZ1}
\end{figure}

\section{Inverted Staircase Algorithm}
\label{sec:inversestaircase}
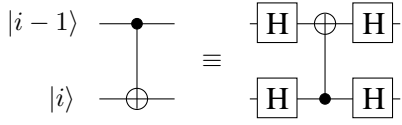
\begin{figure}[h!]
        \centering
        \begin{tikzpicture}
    \draw (6.5,8)--(7.5,8);
    \draw (6.5,7)--(7.5,7);

    \draw (7,7) circle (4pt);
    \filldraw (7,8) circle (2pt);
    \draw (7,8)--(7,6.87);

    \node at (8,7.5){$\equiv$};

    \draw (8.5,8)--(10.5,8);
    \draw (8.5,7)--(10.5,7);

    \filldraw[draw=black, fill=white] (8.6,6.75) rectangle ++(15pt,15pt);
    \node at (8.87,7){\large H};
    \filldraw[draw=black, fill=white] (8.6,7.75) rectangle ++(15pt,15pt);
    \node at (8.87,8){\large H};

    \draw (9.5,8) circle (4pt);
    \filldraw (9.5,7) circle (2pt);
    \draw (9.5,8.13)--(9.5,7);

    \filldraw[draw=black, fill=white] (9.87,6.75) rectangle ++(15pt,15pt);
    \node at (10.15,7){\large H};
    \filldraw[draw=black, fill=white] (9.87,7.75) rectangle ++(15pt,15pt);
    \node at (10.15,8){\large H};

    \node at (5.75,8){$\ket{i-1}$};
    \node at (6,7){$\ket{i}$};

\end{tikzpicture}
	\caption{Equivalence between a standard $\operatorname{CNOT}$ and an inverted one through four Hadamard gates.}
	\label{fig:relationCNOT}
\end{figure}

It is possible to define a similar algorithm to the staircase one which has a better efficiency in the number of one-qubit gates through the existing relation between $\operatorname{CNOT}$ gates and inverted $\operatorname{CNOT}$ gates, 
\begin{equation}
    \operatorname{CNOT}(i,i-1)=(H\otimes H)\operatorname{CNOT}(i-1,i)(H\otimes H)
\end{equation}
 where $i-1$ and $i$ denote the respective qubits, see figure \ref{fig:relationCNOT}.

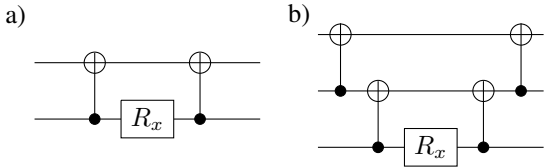
\begin{figure}[h!]
        \centering
        \begin{tikzpicture}

            \filldraw[draw=white, fill=white] (5.15,6.75)--(5.85,6.75)--(5.85,6.25)--(5.15,6.25)--(5.15,6.75); 
            \node at (3.75,8.25){a)};
		
		\draw (4,7.63)--(7,7.63);
            \draw (4,6.88)--(7,6.88);

            \draw (4.8,7.63) circle (4pt);
            \draw (6.2,7.63) circle (4pt);
            \filldraw (4.8,6.88) circle (2pt);
		\filldraw (6.2,6.88) circle (2pt);
            \draw (4.8,7.76)--(4.8,6.88);
		\draw (6.2,7.76)--(6.2,6.88);
		\filldraw[draw=black, fill=white] (5.15,7.13)--(5.85,7.13)--(5.85,6.63)--(5.15,6.63)--(5.15,7.13);
            \node at (5.5,6.88){$R_x$};

	\end{tikzpicture}
        \begin{tikzpicture}
            \node at (3.75,8.25){b)};
		\draw (4,8)--(7,8);
		\draw (4,7.25)--(7,7.25);
            \draw (4,6.5)--(7,6.5);

		\draw (4.3,8) circle (4pt);
		\draw (6.7,8) circle (4pt);
		\filldraw (4.3,7.25) circle (2pt);
		\filldraw (6.7,7.25) circle (2pt);
		\draw (4.3,8.15)--(4.3,7.25);
		\draw (6.7,8.15)--(6.7,7.25);
            \draw (4.8,7.25) circle (4pt);
            \draw (6.2,7.25) circle (4pt);
            \filldraw (4.8,6.5) circle (2pt);
		\filldraw (6.2,6.5) circle (2pt);
            \draw (4.8,7.38)--(4.8,6.5);
		\draw (6.2,7.38)--(6.2,6.5);
		\filldraw[draw=black, fill=white] (5.15,6.75)--(5.85,6.75)--(5.85,6.25)--(5.15,6.25)--(5.15,6.75);
            \node at (5.5,6.5){$R_x$};

	\end{tikzpicture}
	\caption{Quantum circuits depicting how the inverted staircase algorithm works. \textbf{a)} Quantum circuit decomposing the exponential $e^{-ia\sigma_x^1\otimes\sigma_x^2}$. \textbf{b)} Quantum circuit decomposing the exponential $e^{-ia\sigma_x^1\otimes\sigma_x^2\otimes\sigma_x^3}$. The addition of another $\otimes\sigma_x^i$ in the generalized Pauli matrix amounts to extending the quantum circuit with one inverted $\operatorname{CNOT}$ gate at each side.}
	\label{fig:invertedstaircase}
\end{figure}

This \textit{inverted} staircase algorithm starts with the decomposition of the Hamiltonian $H=a\sigma_x^1\otimes\sigma_x^2$ through two \textit{inverted} $\operatorname{CNOT}$ gates surrounding a one-qubit $R_x(2a)$. Then, analogous as before, for every $\sigma_x^i$ it is tensored with, it adds a pair of inverted $\operatorname{CNOT}$ gates, one at each side of the one-qubit gate, see figure \ref{fig:invertedstaircase}. 

Using the same relations (\ref{eq:relation1}) and (\ref{eq:relation2}) as before we can also account for the tensoring of Pauli matrices besides $\sigma_x$. For every $\sigma_z^i$ it is tensored with, it adds a pair of inverted $\operatorname{CNOT}$ gates as well as a pair of $H$ gates, acting on the external side of the $i$th qubit. Similarly, for every $\sigma_y^i$ it is tensored with, it adds a pair of inverted $\operatorname{CNOT}$ gates as well as a pair of one-qubit phase gates, where $R_z\left(\frac{\pi}{2}\right)$ acts on the left side and a $R_z\left(-\frac{\pi}{2}\right)$ on the right side of the $i$th qubit. It is through this relationship that the efficiency in the number of one-qubit gates with respect to the standard staircase algorithm appears.

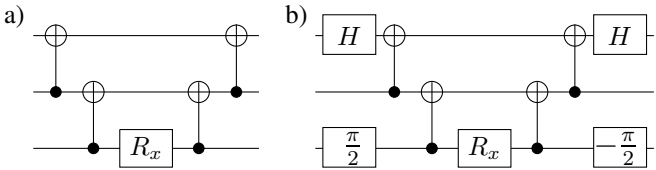
\begin{figure}[h!]
        \centering
        \begin{tikzpicture}
            \node at (3.75,8.25){a)};
		\draw (4,8)--(7,8);
		\draw (4,7.25)--(7,7.25);
            \draw (4,6.5)--(7,6.5);

		\draw (4.3,8) circle (4pt);
		\draw (6.7,8) circle (4pt);
		\filldraw (4.3,7.25) circle (2pt);
		\filldraw (6.7,7.25) circle (2pt);
		\draw (4.3,8.15)--(4.3,7.25);
		\draw (6.7,8.15)--(6.7,7.25);
            \draw (4.8,7.25) circle (4pt);
            \draw (6.2,7.25) circle (4pt);
            \filldraw (4.8,6.5) circle (2pt);
		\filldraw (6.2,6.5) circle (2pt);
            \draw (4.8,7.38)--(4.8,6.5);
		\draw (6.2,7.38)--(6.2,6.5);
		\filldraw[draw=black, fill=white] (5.15,6.75)--(5.85,6.75)--(5.85,6.25)--(5.15,6.25)--(5.15,6.75);
            \node at (5.5,6.5){$R_x$};

            \node at (7.5,8.25){b)};
            \draw (7.75,8)--(12.25,8);
            \draw (7.75,7.25)--(12.25,7.25);
            \draw (7.75,6.5)--(12.25,6.5);
		\filldraw[draw=black, fill=white] (9.65,6.75)--(10.35,6.75)--(10.35,6.25)--(9.65,6.25)--(9.65,6.75);
            \node at (10,6.5){$R_x$};
            \draw (9.3,7.25) circle (4pt);
            \draw (10.7,7.25) circle (4pt);
            \filldraw (9.3,6.5) circle (2pt);
		\filldraw (10.7,6.5) circle (2pt);
            \draw (9.3,7.38)--(9.3,6.5);
		\draw (10.7,7.38)--(10.7,6.5);
            \draw (8.8,8) circle (4pt);
		\draw (11.2,8) circle (4pt);
		\filldraw (8.8,7.25) circle (2pt);
		\filldraw (11.2,7.25) circle (2pt);
		\draw (8.8,8.15)--(8.8,7.25);
		\draw (11.2,8.15)--(11.2,7.25);

            \filldraw[draw=black, fill=white] (7.85,7.75) rectangle ++(20pt,15pt);
            \node at (8.2,8){$H$};
            
            \filldraw[draw=black, fill=white] (7.85,6.25) rectangle ++(20pt,15pt);
            \node at (8.25,6.5){\large $\frac{\pi}{2}$};

            \filldraw[draw=black, fill=white] (11.45,7.75) rectangle ++(20pt,15pt);
            \node at (11.8,8){$H$};
            
            \filldraw[draw=black, fill=white] (11.45,6.25) rectangle ++(20pt,15pt);
            \node at (11.75,6.5){\large $-\frac{\pi}{2}$};

	\end{tikzpicture}
	\caption{Quantum circuits decomposing two exponentials $e^{-iaH}$ showing how the inverted staircase algorithm works. \textbf{a)}$H=\sigma_x^1\otimes\sigma_x^2\otimes\sigma_x^3$. \textbf{b)} $H=\sigma_z^1\otimes\sigma_x^2\otimes\sigma_y^3$, for which it uses the relations (\ref{eq:relation1}) and (\ref{eq:relation2}).}
	\label{fig:XXXvsZXY}
\end{figure}
\begin{figure}[h!]
        \centering
        \begin{tikzpicture}


            
            \draw (7.75,8)--(12.25,8);
            \draw (7.75,7.25)--(12.25,7.25);
            \draw (7.75,6.5)--(12.25,6.5);
		\filldraw[draw=black, fill=white] (9.65,6.75)--(10.35,6.75)--(10.35,6.25)--(9.65,6.25)--(9.65,6.75);
            \node at (10,6.5){$R_x$};
            \draw (9.3,7.25) circle (4pt);
            \draw (10.7,7.25) circle (4pt);
            \filldraw (9.3,6.5) circle (2pt);
		\filldraw (10.7,6.5) circle (2pt);
            \draw (9.3,7.38)--(9.3,6.5);
		\draw (10.7,7.38)--(10.7,6.5);
            \draw (8.8,8) circle (4pt);
		\draw (11.2,8) circle (4pt);
		\filldraw (8.8,7.25) circle (2pt);
		\filldraw (11.2,7.25) circle (2pt);
		\draw (8.8,8.15)--(8.8,7.25);
		\draw (11.2,8.15)--(11.2,7.25);

            \filldraw[draw=black, fill=white] (7.85,7.75) rectangle ++(20pt,15pt);
            \node at (8.25,8){\large $\frac{\pi}{2}$};
            \filldraw[draw=black, fill=white] (7.85,7) rectangle ++(20pt,15pt);
            \node at (8.25,7.25){\large $\frac{\pi}{2}$};
            \filldraw[draw=black, fill=white] (7.85,6.25) rectangle ++(20pt,15pt);
            \node at (8.25,6.5){\large $\frac{\pi}{2}$};

            \filldraw[draw=black, fill=white] (11.45,7.75) rectangle ++(20pt,15pt);
            \node at (11.75,8){\large $-\frac{\pi}{2}$};
            \filldraw[draw=black, fill=white] (11.45,7) rectangle ++(20pt,15pt);
            \node at (11.75,7.25){\large $-\frac{\pi}{2}$};
            \filldraw[draw=black, fill=white] (11.45,6.25) rectangle ++(20pt,15pt);
            \node at (11.75,6.5){\large $-\frac{\pi}{2}$};

	\end{tikzpicture}
	\caption{Quantum circuit decomposing the exponential $e^{-ia\sigma_y^1\otimes\sigma_y^2\otimes\sigma_y^3}$ through the inverted staircase algorithm. The difference between this quantum circuit and the quantum circuit simulating $H=\sigma_x^1\otimes\sigma_x^2\otimes\sigma_x^3$ is in the placement of external phases which depend on the position of the Pauli matrices $\sigma_y^i$.}
	\label{fig:XXXvsYYY}
\end{figure}
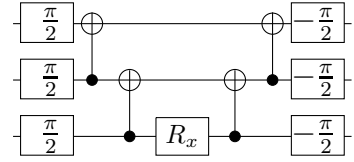

\begin{remark}
Although it might seem that the use of the rotation matrix $R_y$ to simulate the exponential of the generalized Pauli matrix $H=-iaYYY$ can be done only through inverted $\operatorname{CNOT}$ gates and without using one-qubit gates, it turns out that it does not work. Thus, the most efficient-staircase way is through the rotation matrix $R_x$ with six inverted $\operatorname{CNOT}$ gates and eight one-qubit phase gates, see figure \ref{fig:XXXvsYYY}.
\end{remark}

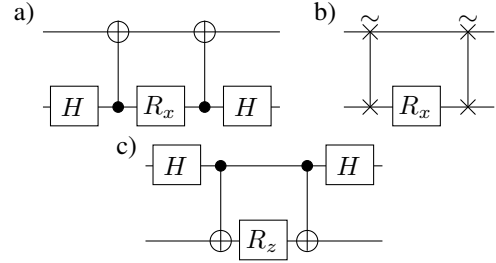
\begin{figure}[h!]
        \centering
        \begin{tikzpicture}
            \node at (3.5,8.25){a)};
		\draw (3.75,8)--(6.9,8);
		\draw (3.75,7)--(6.9,7);

		\draw (4.75,8) circle (4pt);
		\draw (5.9,8) circle (4pt);
		\filldraw (4.75,7) circle (2pt);
		\filldraw (5.9,7) circle (2pt);
		\draw (4.75,8.15)--(4.75,7);
		\draw (5.9,8.15)--(5.9,7);
            \filldraw[draw=black, fill=white] (5,6.75) rectangle ++(18pt,15pt);
            \node at (5.3,7){$R_x$};
            \filldraw[draw=black, fill=white] (3.85,6.75) rectangle ++(18pt,15pt);
            \node at (4.15,7){$H$};
            \filldraw[draw=black, fill=white] (6.15,6.75) rectangle ++(18pt,15pt);
            \node at (6.45,7){$H$};

            \node at (7.5,8.25){b)};
            \draw (7.75,8)--(9.75,8);
            \draw (7.75,7)--(9.75,7);
            \filldraw[draw=black, fill=white] (8.4,6.75) rectangle ++(18pt,15pt);
            \node at (8.7,7){$R_x$};

            \node at (8.1,8.15){$\sim$};
            \draw (8.1,8)--(8.1,7);
            \draw (8,7.9)--(8.2,8.1);
            \draw (8,8.1)--(8.2,7.9);
            \draw (8,6.9)--(8.2,7.1);
            \draw (8,7.1)--(8.2,6.9);
            \node at (9.4,8.15){$\sim$};
            \draw (9.4,8)--(9.4,7);	
            \draw (9.3,7.9)--(9.5,8.1);
            \draw (9.3,8.1)--(9.5,7.9);
            \draw (9.3,6.9)--(9.5,7.1);
            \draw (9.3,7.1)--(9.5,6.9);
	\end{tikzpicture}
        \begin{tikzpicture}
            \node at (3.5,8.25){c)};
		\draw (3.75,8)--(6.9,8);
		\draw (3.75,7)--(6.9,7);

		\draw (4.75,7) circle (4pt);
		\draw (5.9,7) circle (4pt);
		\filldraw (4.75,8) circle (2pt);
		\filldraw (5.9,8) circle (2pt);
		\draw (4.75,8)--(4.75,6.87);
		\draw (5.9,8)--(5.9,6.87);
            \filldraw[draw=black, fill=white] (5,6.75) rectangle ++(18pt,15pt);
            \node at (5.3,7){$R_z$};
            \filldraw[draw=black, fill=white] (3.85,7.75) rectangle ++(18pt,15pt);
            \node at (4.15,8){$H$};
            \filldraw[draw=black, fill=white] (6.15,7.75) rectangle ++(18pt,15pt);
            \node at (6.45,8){$H$};		
	\end{tikzpicture}
	\caption{Realizations of the exponential $e^{-ia \sigma_x^1\otimes\sigma_z^2}$ through \textbf{a)} Inverted staircase algorithm. \textbf{b)} Through fermionic $\operatorname{SWAP}$ gates. \textbf{c)} Staircase algorithm.}
	\label{fig:XZ}
\end{figure}

It is relatively straightforward to transpose the staircase algorithm into the inverted one. This can be shown by replacing each $\operatorname{CNOT}$ gate with an inverted $\operatorname{CNOT}$ gate with four surrounding Hadamard gates, see figure \ref{fig:relationCNOT}, and accounting for the relations $HH=\mathbbm{1}$ and $HR_xH=R_z$. In figure \ref{fig:XZ} we have included three different quantum circuits simulating the dynamics of the Hamiltonian $H=a\sigma_x\otimes\sigma_z$.

\section{Expansion of the Inverted Staircase Algorithm through fermionic gates}
\label{sec:algorithm}

Through the inverted staircase algorithm, it is possible to define an equivalent algorithm simulating the exponential of a generalized Pauli matrix using fermionic $\operatorname{SWAP}$ gates. If implementable, such fermionic quantum gates allow a reduction in the number of $\operatorname{CNOT}$ gates used to simulate the exponential of a Hamiltonian involving generalized Pauli matrices and offer a simplification of quantum gates in a variety of quantum circuits.

Suppose we have a Hamiltonian involving only one generalized Pauli basis acting on an $n$-qubit system,
\begin{equation}
    H=a\sigma_{i}^1\otimes\sigma_{i}^2\otimes\cdots\otimes\sigma_{i}^{n-1}\otimes\sigma_{i}^n
\end{equation}
where $\sigma_i, i\in[0,1,2,3]$, denote the standard Pauli matrices spanning $SU(2)$ and $a$ is an arbitrary constant. As before, in order to simulate $e^{-iH}$, we start by decomposing the exponential of the first two terms of the Hamiltonian $\sigma_{i}^1\otimes\sigma_{i}^2$ and build up through an algorithm that recursively expands the quantum circuit by systematically adding $\operatorname{CNOT}$, $\operatorname{SWAP}$, and $\operatorname{fSWAP}$ gates according to the following rules: 
\begin{enumerate}
	\item If the exponential is enlarged through $\otimes\sigma_x^3$, or equivalently through $\otimes\sigma_y^3$, then two inverted $\operatorname{CNOT}$ gates are added at the nearest side of the central rotation $R_x$. As for the inverted staircase algorithm, for an expansion through $\otimes\sigma_y$ it is necessary to add one-qubit phase gates at the external sites, see figure \ref{fig:XXY}.
 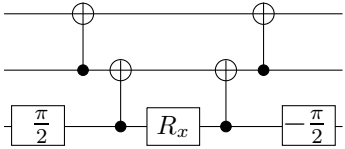
\begin{figure}[h!]
        \centering
	\begin{tikzpicture}
		
            \draw (7.75,8)--(12.25,8);
            \draw (7.75,7.25)--(12.25,7.25);
            \draw (7.75,6.5)--(12.25,6.5);
		\filldraw[draw=black, fill=white] (9.65,6.75)--(10.35,6.75)--(10.35,6.25)--(9.65,6.25)--(9.65,6.75);
            \node at (10,6.5){$R_x$};
            \draw (9.3,7.25) circle (4pt);
            \draw (10.7,7.25) circle (4pt);
            \filldraw (9.3,6.5) circle (2pt);
		\filldraw (10.7,6.5) circle (2pt);
            \draw (9.3,7.38)--(9.3,6.5);
		\draw (10.7,7.38)--(10.7,6.5);
            \draw (8.8,8) circle (4pt);
		\draw (11.2,8) circle (4pt);
		\filldraw (8.8,7.25) circle (2pt);
		\filldraw (11.2,7.25) circle (2pt);
		\draw (8.8,8.15)--(8.8,7.25);
		\draw (11.2,8.15)--(11.2,7.25);

            \filldraw[draw=black, fill=white] (7.85,6.25) rectangle ++(20pt,15pt);
            \node at (8.25,6.5){\large $\frac{\pi}{2}$};

            \filldraw[draw=black, fill=white] (11.45,6.25) rectangle ++(20pt,15pt);
            \node at (11.75,6.5){\large $-\frac{\pi}{2}$};

	\end{tikzpicture}	
	\caption{We obtain the quantum circuit decomposing the exponential $e^{-ia\sigma_x^1\otimes\sigma_x^2\otimes\sigma_y^3}$ through enlarging the quantum circuit decomposing the exponential $e^{-ia\sigma_x^1\otimes\sigma_x^2}$ with two $\operatorname{CNOT}$ gates, one at each side, and the respective phases corresponding to be tensoring with a $\sigma_y^3$ matrix.}
	\label{fig:XXY}
\end{figure}
	\item If the exponential is enlarged through $\otimes\sigma_z^3$, then two $\operatorname{fSWAP}$ gates are added at the nearest side of the central rotation $R_x$, see figure \ref{fig:XXZ}. 
  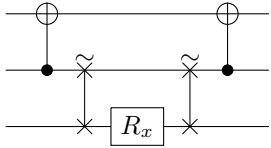
\begin{figure}[h!]
        \centering
	\begin{tikzpicture}
		
            \draw (8.25,8)--(11.75,8);
            \draw (8.25,7.25)--(11.75,7.25);
            \draw (8.25,6.5)--(11.75,6.5);
		\filldraw[draw=black, fill=white] (9.65,6.75)--(10.35,6.75)--(10.35,6.25)--(9.65,6.25)--(9.65,6.75);
            \node at (10,6.5){$R_x$};

            \node at (9.3,7.4){$\sim$};            
            \draw (9.2,7.15)--(9.4,7.35);
            \draw (9.2,7.35)--(9.4,7.15);
            \draw (9.2,6.4)--(9.4,6.6);
            \draw (9.2,6.6)--(9.4,6.4);
            \node at (10.7,7.4){$\sim$}; 
            \draw (10.6,7.15)--(10.8,7.35);
            \draw (10.6,7.35)--(10.8,7.15);
            \draw (10.6,6.4)--(10.8,6.6);
            \draw (10.6,6.6)--(10.8,6.4);
            
            \draw (9.3,7.25)--(9.3,6.5);
		\draw (10.7,7.25)--(10.7,6.5);
  
            \draw (8.8,8) circle (4pt);
		\draw (11.2,8) circle (4pt);
		\filldraw (8.8,7.25) circle (2pt);
		\filldraw (11.2,7.25) circle (2pt);
		\draw (8.8,8.15)--(8.8,7.25);
		\draw (11.2,8.15)--(11.2,7.25);

	\end{tikzpicture}	
	\caption{We obtain the quantum circuit decomposing the exponential $e^{-ia\sigma_x^1\otimes\sigma_x^2\otimes\sigma_z^3}$ through enlarging the quantum circuit decomposing the exponential $e^{-ia\sigma_x^1\otimes\sigma_x^2}$ with two $\operatorname{fSWAP}$ gates, one at each side.}
	\label{fig:XXZ}
\end{figure}
\item If the exponential is enlarged through $\otimes\mathbbm{1}^3$, then two $\operatorname{SWAP}$ gates are added at the nearest side of the central rotation $R_x$. As before, this is equivalent to adding an unaffected qubit in the position where the $\mathbbm{1}$ matrices are in the Hamiltonian.
\end{enumerate}

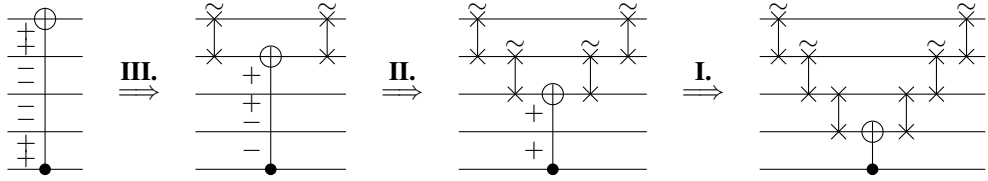
\begin{figure*}[h]
        \centering
        \begin{tikzpicture}

\draw (5,6)--(6,6);
\draw (5,5.5)--(6,5.5);
\draw (5,5)--(6,5);
\draw (5,4.5)--(6,4.5);
\draw (5,4)--(6,4);

\draw (5.5,6) circle (4pt);
\filldraw (5.5,4) circle (2pt);
\draw (5.5,6.15)--(5.5,4);
\node at (5.25,5.83){$+$};
\node at (5.25,5.63){$+$};
\node at (5.25,5.34){$-$};
\node at (5.25,5.16){$-$};
\node at (5.25,4.84){$-$};
\node at (5.25,4.66){$-$};
\node at (5.25,4.35){$+$};
\node at (5.25,4.15){$+$};

\node at (6.75,5){$\implies$};	
\node at (6.75,5.3){\textbf{III.}};

\draw (7.5,6)--(9.5,6);
\draw (7.5,5.5)--(9.5,5.5);
\draw (7.5,5)--(9.5,5);
\draw (7.5,4.5)--(9.5,4.5);
\draw (7.5,4)--(9.5,4);

\draw (8.5,5.5) circle (4pt);
\filldraw (8.5,4) circle (2pt);
\draw (8.5,5.63)--(8.5,4);
\node at (8.25,5.25){$+$};
\node at (8.25,4.88){$+$};
\node at (8.25,4.63){$-$};
\node at (8.25,4.25){$-$};

\node at (7.75,6.15){$\sim$};
\draw (7.75,6)--(7.75,5.5);
\draw (7.65,5.9)--(7.85,6.1);
\draw (7.65,6.1)--(7.85,5.9);
\draw (7.65,5.4)--(7.85,5.6);
\draw (7.65,5.6)--(7.85,5.4);
\node at (9.25,6.15){$\sim$};
\draw (9.25,6)--(9.25,5.5);
\draw (9.15,5.9)--(9.35,6.1);
\draw (9.15,6.1)--(9.35,5.9);
\draw (9.15,5.4)--(9.35,5.6);
\draw (9.15,5.6)--(9.35,5.4);

\node at (10.25,5){$\implies$};	
\node at (10.25,5.3){\textbf{II.}};

\draw (11,6)--(13.5,6);
\draw (11,5.5)--(13.5,5.5);
\draw (11,5)--(13.5,5);
\draw (11,4.5)--(13.5,4.5);
\draw (11,4)--(13.5,4);

\draw (12.25,5) circle (4pt);
\filldraw (12.25,4) circle (2pt);
\draw (12.25,5.13)--(12.25,4);
\node at (12,4.75){$+$};
\node at (12,4.25){$+$};

\node at (11.25,6.15){$\sim$};
\draw (11.25,6)--(11.25,5.5);
\draw (11.15,5.9)--(11.35,6.1);
\draw (11.15,6.1)--(11.35,5.9);
\draw (11.15,5.4)--(11.35,5.6);
\draw (11.15,5.6)--(11.35,5.4);
\node at (13.25,6.15){$\sim$};
\draw (13.25,6)--(13.25,5.5);
\draw (13.15,5.9)--(13.35,6.1);
\draw (13.15,6.1)--(13.35,5.9);
\draw (13.15,5.4)--(13.35,5.6);
\draw (13.15,5.6)--(13.35,5.4);

\node at (11.75,5.65){$\sim$};
\draw (11.75,5.5)--(11.75,5);
\draw (11.65,5.4)--(11.85,5.6);
\draw (11.65,5.6)--(11.85,5.4);
\draw (11.65,4.9)--(11.85,5.1);
\draw (11.65,5.1)--(11.85,4.9);
\node at (12.75,5.65){$\sim$};
\draw (12.75,5.5)--(12.75,5);
\draw (12.65,5.4)--(12.85,5.6);
\draw (12.65,5.6)--(12.85,5.4);
\draw (12.65,4.9)--(12.85,5.1);
\draw (12.65,5.1)--(12.85,4.9);

\node at (14.25,5){$\implies$};	
\node at (14.25,5.3){\textbf{I.}};

\draw (15,6)--(18,6);
\draw (15,5.5)--(18,5.5);
\draw (15,5)--(18,5);
\draw (15,4.5)--(18,4.5);
\draw (15,4)--(18,4);

\draw (16.5,4.5) circle (4pt);
\filldraw (16.5,4) circle (2pt);
\draw (16.5,4.63)--(16.5,4);

\node at (15.25,6.15){$\sim$};
\draw (15.25,6)--(15.25,5.5);
\draw (15.15,5.9)--(15.35,6.1);
\draw (15.15,6.1)--(15.35,5.9);
\draw (15.15,5.4)--(15.35,5.6);
\draw (15.15,5.6)--(15.35,5.4);
\node at (17.75,6.15){$\sim$};
\draw (17.75,6)--(17.75,5.5);
\draw (17.65,5.9)--(17.85,6.1);
\draw (17.65,6.1)--(17.85,5.9);
\draw (17.65,5.4)--(17.85,5.6);
\draw (17.65,5.6)--(17.85,5.4);

\node at (15.65,5.65){$\sim$};
\draw (15.65,5.5)--(15.65,5);
\draw (15.55,5.4)--(15.75,5.6);
\draw (15.55,5.6)--(15.75,5.4);
\draw (15.55,4.9)--(15.75,5.1);
\draw (15.55,5.1)--(15.75,4.9);
\node at (17.35,5.65){$\sim$};
\draw (17.35,5.5)--(17.35,5);
\draw (17.25,5.4)--(17.45,5.6);
\draw (17.25,5.6)--(17.45,5.4);
\draw (17.25,4.9)--(17.45,5.1);
\draw (17.25,5.1)--(17.45,4.9);

\draw (16.05,5)--(16.05,4.5);
\draw (15.95,4.9)--(16.15,5.1);
\draw (15.95,5.1)--(16.15,4.9);
\draw (15.95,4.4)--(16.15,4.6);
\draw (15.95,4.6)--(16.15,4.4);
\draw (16.95,5)--(16.95,4.5);
\draw (16.85,4.9)--(17.05,5.1);
\draw (16.85,5.1)--(17.05,4.9);
\draw (16.85,4.4)--(17.05,4.6);
\draw (16.85,4.6)--(17.05,4.4);
\end{tikzpicture}
	\caption{Keeping track of the position of the negative signs in the dominant entries of the off-diagonal allows us to construct fermionic $\operatorname{CNOT}$ gates. Such gates, if implementable on quantum circuits, improve the efficiency of the standard staircase algorithm.}
	\label{fig:CNOT(++----++)}
\end{figure*}

This construction also generalizes to the $n$-fold product of $\sigma_i$ matrices and  allows the merging of non-directly connected fermionic quantum gates to separate fermionic quantum gates. For instance, through the relations explained in section \ref{sec:gates}, we can enlarge a $\operatorname{SWAP}(n-1,n)$ or $\operatorname{CNOT}(n,n-1)$ quantum gate by using $\operatorname{SWAP}$ or $\operatorname{fSWAP}$ gates to a $\operatorname{SWAP}(n-2,n)$ or $\operatorname{CNOT}(n,n-2)$ quantum gate. An example of an enlarged fermionic $\operatorname{CNOT}(5,1)$ gate can be seen in figure \ref{fig:CNOT(++----++)}.

\section{Examples}
\label{sec:examples}
\begin{figure*}[h]
    \centering
        \begin{tikzpicture}
		\draw (5.05,5.5)--(6.75,5.5);
            \draw (5.0,4.5)--(6.75,4.5);
            \filldraw[draw=black, fill=white] (5.5,4.75)--(6.3,4.75)--(6.3,4.25)--(5.5,4.25)--(5.5,4.75);
            \node at (5.85,4.5){$X$};

    \node at (7.5,5){$\implies$};	
    \node at (7.5,5.3){\textbf{I.}};
    \node at (7.5,4.7){$\otimes\sigma_x$};

            \draw (8.25,5.75)--(10.75,5.75);
            \draw (8.25,5)--(10.75,5);
            \draw (8.25,4.25)--(10.75,4.25);
            \filldraw[draw=black, fill=white] (9.15,4.5)--(9.85,4.5)--(9.85,4)--(9.15,4)--(9.15,4.5);
            \node at (9.5,4.25){$R_x$};          

            \draw (8.6,5) circle (4pt);
            \draw (8.6,5.13)--(8.6,4.25);
            \filldraw (8.6,4.25) circle (2pt);
            
            \draw (10.4,5) circle (4pt);
            \draw (10.4,5.13)--(10.4,4.25);
            \filldraw (10.4,4.25) circle (2pt);

    \node at (11.5,5){$\implies$};	
    \node at (11.5,5.3){\textbf{II.}};
    \node at (11.5,4.7){$\otimes\mathbbm{1}$};

            \draw (12.25,5.75)--(15.25,5.75);
            \draw (12.25,5.25)--(15.25,5.25);
            \draw (12.25,4.75)--(15.25,4.75);
            \draw (12.25,4.25)--(15.25,4.25);
            \filldraw[draw=black, fill=white] (13.4,4.5)--(14.1,4.5)--(14.1,4)--(13.4,4)--(13.4,4.5);
            \node at (13.75,4.25){$R_x$};

            \draw (12.55,5.25) circle (4pt);
            \draw (12.55,5.38)--(12.55,4.75);
            \filldraw (12.55,4.75) circle (2pt);

            \draw (13,4.75)--(13,4.25);
            \draw (12.9,4.65)--(13.1,4.85);
            \draw (12.9,4.85)--(13.1,4.65);
            \draw (12.9,4.15)--(13.1,4.35);
            \draw (12.9,4.35)--(13.1,4.15);

            \draw (14.95,5.25) circle (4pt);
            \draw (14.95,5.38)--(14.95,4.75);
            \filldraw (14.95,4.75) circle (2pt);

            \draw (14.5,4.75)--(14.5,4.25);
            \draw (14.4,4.65)--(14.6,4.85);
            \draw (14.4,4.85)--(14.6,4.65);
            \draw (14.4,4.15)--(14.6,4.35);
            \draw (14.4,4.35)--(14.6,4.15);
    \node at (16,5){$\implies$};	
    \node at (16,5.3){\textbf{III.}};
    \node at (16,4.7){$\otimes\sigma_z$};

            \draw (16.75,6)--(20.25,6);
            \draw (16.75,5.5)--(20.25,5.5);
            \draw (16.75,5)--(20.25,5);
            \draw (16.75,4.5)--(20.25,4.5);
            \draw (16.75,4)--(20.25,4);
            \filldraw[draw=black, fill=white] (18.1,4.25)--(18.9,4.25)--(18.9,3.75)--(18.1,3.75)--(18.1,4.25);
            \node at (18.5,4){$R_x$};

            \draw (17.05,5.5) circle (4pt);
            \draw (17.05,5.63)--(17.05,5);
            \filldraw (17.05,5) circle (2pt);
            \draw (19.95,5.5) circle (4pt);
            \draw (19.95,5.63)--(19.95,5);
            \filldraw (19.95,5) circle (2pt);

            \draw (17.5,5)--(17.5,4);
            \node at (17.5,5.15){$\sim$};
            \draw (17.4,4.9)--(17.6,5.1);
            \draw (17.4,5.1)--(17.6,4.9);
            \draw (17.4,3.9)--(17.6,4.1);
            \draw (17.4,4.1)--(17.6,3.9);

            \node at (17.35,4.75){$+$};
            \node at (17.35,4.25){$+$};

            \draw (19.5,5)--(19.5,4);
            \node at (19.5,5.15){$\sim$};
            \draw (19.4,4.9)--(19.6,5.1);
            \draw (19.4,5.1)--(19.6,4.9);
            \draw (19.4,3.9)--(19.6,4.1);
            \draw (19.4,4.1)--(19.6,3.9);

            \node at (19.35,4.75){$+$};
            \node at (19.35,4.25){$+$};
            
\end{tikzpicture}
	\caption{\textbf{Example 1: }Quantum circuit decomposing the exponential $e^{-i\sigma_x^2\sigma_x^3\sigma_z^5}$. \textbf{I.} Through enlarging with a $\sigma_x^3$ matrix we obtain the same case as with the base case $e^{-i\sigma_x\otimes\sigma_x}$, where the first qubit remains unaffected. \textbf{II.} We expand the quantum circuit by adding two $\operatorname{SWAP}(3,4)$ gates. \textbf{III.} In the last step we add two $\operatorname{SWAP}(4,5)$ gates, which are simplified by the classical $\operatorname{SWAP}$ gates and become $\operatorname{fSWAP}(3,5)$. Note that the construction of the quantum circuit decomposing the exponential $e^{-i\sigma_y^2\sigma_y^3\sigma_z^5}$ is analogous with the respective phase gates at the external side of the second and third qubits.}
	\label{fig:1XX1Z}
\end{figure*}
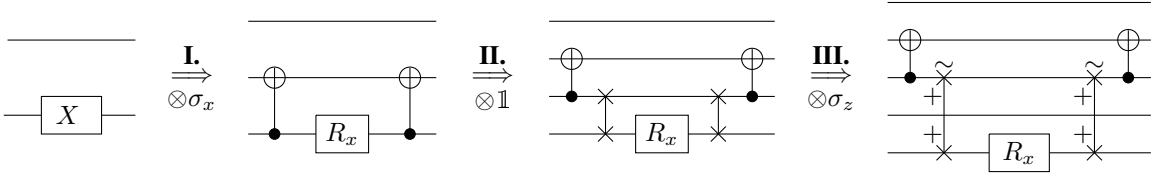

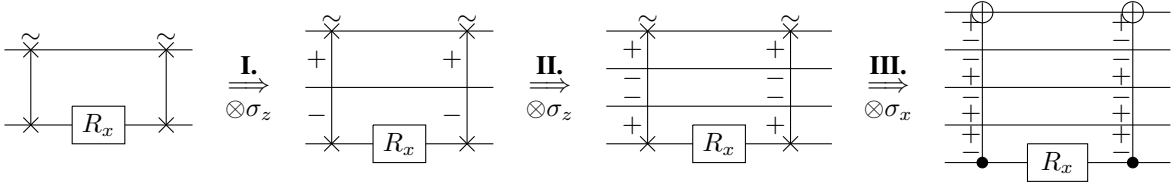
\begin{figure*}[h]
        \centering
        \begin{tikzpicture}
		\draw (4.25,5.5)--(6.75,5.5);
            \draw (4.25,4.5)--(6.75,4.5);
            \filldraw[draw=black, fill=white] (5.15,4.75)--(5.85,4.75)--(5.85,4.25)--(5.15,4.25)--(5.15,4.75);
            \node at (5.5,4.5){$R_x$};
            
            \node at (4.6,5.65){$\sim$};
            \draw (4.5,5.4)--(4.7,5.6);
            \draw (4.5,5.6)--(4.7,5.4);
            \draw (4.5,4.4)--(4.7,4.6);
            \draw (4.5,4.6)--(4.7,4.4);
            \draw (4.6,5.5)--(4.6,4.5);

            \node at (6.4,5.65){$\sim$};
            \draw (6.3,5.4)--(6.5,5.6);
            \draw (6.3,5.6)--(6.5,5.4);
            \draw (6.3,4.4)--(6.5,4.6);
            \draw (6.3,4.6)--(6.5,4.4);
            \draw (6.4,5.5)--(6.4,4.5);

    \node at (7.5,5){$\implies$};	
    \node at (7.5,5.3){\textbf{I.}};
    \node at (7.5,4.7){$\otimes\sigma_z$};

            \draw (8.25,5.75)--(10.75,5.75);
            \draw (8.25,5)--(10.75,5);
            \draw (8.25,4.25)--(10.75,4.25);
            \filldraw[draw=black, fill=white] (9.15,4.5)--(9.85,4.5)--(9.85,4)--(9.15,4)--(9.15,4.5);
            \node at (9.5,4.25){$R_x$};

            \node at (8.6,5.9){$\sim$};
            \node at (8.4,5.37){$+$};
            \node at (8.4,4.63){$-$};
            \draw (8.5,5.65)--(8.7,5.85);
            \draw (8.5,5.85)--(8.7,5.65);
            \draw (8.5,4.15)--(8.7,4.35);
            \draw (8.5,4.35)--(8.7,4.15);
            \draw (8.6,5.75)--(8.6,4.25);

            \node at (10.4,5.9){$\sim$};
            \node at (10.2,5.37){$+$};
            \node at (10.2,4.63){$-$};
            \draw (10.3,5.65)--(10.5,5.85);
            \draw (10.3,5.85)--(10.5,5.65);
            \draw (10.3,4.15)--(10.5,4.35);
            \draw (10.3,4.35)--(10.5,4.15);
            \draw (10.4,5.75)--(10.4,4.25);

    \node at (11.5,5){$\implies$};	
    \node at (11.5,5.3){\textbf{II.}};
    \node at (11.5,4.7){$\otimes\sigma_z$};

            \draw (12.25,5.75)--(15.25,5.75);
            \draw (12.25,5.25)--(15.25,5.25);
            \draw (12.25,4.75)--(15.25,4.75);
            \draw (12.25,4.25)--(15.25,4.25);
            \filldraw[draw=black, fill=white] (13.4,4.5)--(14.1,4.5)--(14.1,4)--(13.4,4)--(13.4,4.5);
            \node at (13.75,4.25){$R_x$};

            \node at (12.8,5.9){$\sim$};
            \node at (12.6,5.5){$+$};
            \node at (12.6,5.12){$-$};
            \node at (12.6,4.87){$-$};
            \node at (12.6,4.5){$+$};
            \draw (12.7,5.65)--(12.9,5.85);
            \draw (12.7,5.85)--(12.9,5.65);
            \draw (12.7,4.15)--(12.9,4.35);
            \draw (12.7,4.35)--(12.9,4.15);
            \draw (12.8,5.75)--(12.8,4.25);

            \node at (14.7,5.9){$\sim$};
            \node at (14.5,5.5){$+$};
            \node at (14.5,5.12){$-$};
            \node at (14.5,4.87){$-$};
            \node at (14.5,4.5){$+$};
            \draw (14.6,5.65)--(14.8,5.85);
            \draw (14.6,5.85)--(14.8,5.65);
            \draw (14.6,4.15)--(14.8,4.35);
            \draw (14.6,4.35)--(14.8,4.15);
            \draw (14.7,5.75)--(14.7,4.25);

    \node at (16,5){$\implies$};	
    \node at (16,5.3){\textbf{III.}};
    \node at (16,4.7){$\otimes\sigma_x$};

            \draw (16.75,6)--(19.75,6);
            \draw (16.75,5.5)--(19.75,5.5);
            \draw (16.75,5)--(19.75,5);
            \draw (16.75,4.5)--(19.75,4.5);
            \draw (16.75,4)--(19.75,4);
            \filldraw[draw=black, fill=white] (17.85,4.25)--(18.65,4.25)--(18.65,3.75)--(17.85,3.75)--(17.85,4.25);
            \node at (18.25,4){$R_x$};

            \node at (17.1,5.87){$+$};
            \node at (17.1,5.63){$-$};
            \node at (17.1,5.37){$-$};
            \node at (17.1,5.15){$+$};
            \node at (17.1,4.87){$-$};
            \node at (17.1,4.65){$+$};
            \node at (17.1,4.37){$+$};
            \node at (17.1,4.13){$-$};
            \draw (17.25,6) circle (4pt);
            \draw (17.25,6.13)--(17.25,4);
            \filldraw (17.25,4) circle (2pt);

            \node at (19.1,5.87){$+$};
            \node at (19.1,5.63){$-$};
            \node at (19.1,5.37){$-$};
            \node at (19.1,5.15){$+$};
            \node at (19.1,4.87){$-$};
            \node at (19.1,4.65){$+$};
            \node at (19.1,4.37){$+$};
            \node at (19.1,4.13){$-$};
            \draw (19.25,6) circle (4pt);
            \draw (19.25,6.13)--(19.25,4);
            \filldraw (19.25,4) circle (2pt);
\end{tikzpicture}
	\caption{\textbf{Example 2: }Quantum circuit decomposing the exponential $e^{-i\sigma_x^1\sigma_z^2\sigma_z^3\sigma_z^4\sigma_x^5}$. \textbf{I.} Through tensoring with a $\sigma_z^3$ matrix, we expand the previous quantum circuit by adding two $\operatorname{fSWAP}(2,3)$ gates at the inner side. We simplify the quantum circuit by unifying the $\operatorname{fSWAP}(1,2)$ and $\operatorname{fSWAP}(2,3)$ to an $\operatorname{fSWAP}(1,3)$ quantum gate. \textbf{II.} We expand the quantum circuit by adding two $\operatorname{fSWAP}(3,4)$ gates, which simplify through the $\operatorname{fSWAP}(1,3)$ to give two $\operatorname{fSWAP}(1,4)$ quantum gates. \textbf{III.} Through tensoring with $\sigma_x^5$, we expand the quantum circuit by adding two $\operatorname{CNOT}(5,4)$ gates at the inner side. Such a quantum circuit decomposition, if implementable, is more efficient than the standard staircase algorithm since it requires six $\operatorname{CNOT}$ gates and four one-qubit gates less.}
	\label{fig:XZZZX}
\end{figure*}

To illustrate  how the expansion of the inverted staircase algorithm through fermionic gates works, we have included the quantum circuits simulating some randomly-generated Hamiltonians, see figures \ref{fig:1XX1Z} and \ref{fig:XZZZX}. From these examples, it is clear that fermionic gates, if implementable, reduce the number of CNOT gates required in the inverted staircase algorithm. 

\section{Comparison between the three algorithms}
\label{sec:comparison}
It is clear that the inverted staircase algorithm requires the same number of $\operatorname{CNOT}$ gates as the standard staircase algorithm. However, when considering one-qubit gates, there are some generalized Pauli matrices for which it is more efficient to use the staircase algorithm and others for which  the inverted staircase algorithm is more efficient. As a general rule, if the number of $\sigma_z$ in the generalized Pauli matrix is bigger than the number of $\sigma_{\sigma_x}$ plus the number of $\sigma_{\sigma_y}$,
\begin{equation}
    N_{\sigma_z}> N_{\sigma_x}+N_{\sigma_y}
\end{equation}
then it is more efficient, with respect to the number of one-qubit gates, to use the staircase algorithm. Otherwise, that is if the number of $\sigma_z$ is smaller than the number of $\sigma_x$ plus the number of $\sigma_y$,
\begin{equation}
    N_{\sigma_z}< N_{\sigma_x}+N_{\sigma_y}
\end{equation}
it is more efficient to use the inverted staircase algorithm. In case it is equal, that is
\begin{equation}
    N_{\sigma_z}= N_{\sigma_x}+N_{\sigma_y}
\end{equation}
then the one-qubit gates being used in both algorithms are the same.

Assuming that all the Pauli matrices $\sigma_i$, $i\in[0,1,2,3]$, appear with the same probability, in most cases it will be more efficient to use the inverted staircase algorithm. However, the difference between the number of one-qubit gates between the standard and the inverted staircase algorithm is polynomial. This can be shown by determining that in the standard staircase algorithm, the number of one-qubit gates scales on average as
\begin{equation*}
    N_{\text{one-qubit}}^{\text{st}}=2N_{\sigma_x}+4N_{\sigma_y}=2\frac{n}{4}+4\frac{n}{4}=\frac{3n}{2}
\end{equation*}
where $n$ denotes the number of Pauli matrices $\sigma_i\in SU(2)$ in $SU(2^n)$. In the inverted staircase algorithm, the number of one-qubit gates scales on average as
\begin{equation*}
    N_{\text{one-qubit}}^{\text{inv}}=2N_{\sigma_y}+2N_{\sigma_z}=2\frac{n}{4}+2\frac{n}{4}=n
\end{equation*}
Thus, the difference between the number of one-qubit gates in both algorithms scales as
\begin{equation}
    \Delta N_{\text{one-qubit}}=\left|\frac{3n}{2}-n\right|=\frac{n}{2}
\end{equation}

Furthermore, the use of fermionic gates reduces the required number of $\operatorname{CNOT}$ quantum gates to simulate the exponential of a generalized Pauli matrix in comparison to both staircase algorithms by a polynomial factor. As a first approximation, this can be seen by assuming that each of the Pauli matrices appears with the same probability so that for each $\sigma_z^i$, $i\in[2,n-1]$, in the Hamiltonian we spare two $\operatorname{CNOT}$ gates, that is
\begin{equation}
    \Delta N_{\operatorname{CNOT}}\sim 2\frac{n-2}{4}\sim\frac{n}{2}
\end{equation}
where, as before, $n$ denotes the number of Pauli matrices and in the last step we have assumed $n\gg 1$. Note, however, that this is not exact since, for instance, it does not hold for the case in which there are no $\sigma_x$ or $\sigma_y$ in the Hamiltonian. Furthermore, by using fermionic gates there is no improvement in the number of one-qubit gates in comparison to the inverted staircase algorithm.

\section{Suzuki-Trotter Decomposition}
\label{sec:trotter}

In general, the terms composing a Hamiltonian do not commute with each other. Thus, to simulate the exponential of a Hamiltonian through a quantum circuit it is necessary to find an approximation that allows decomposing the exponential. A widely used decomposition for such exponential terms is the Suzuki-Trotter decomposition, \cite{whitfield2011simulation}, \cite{hatano2005suzuki}. This approximation allows simulating the exponential of an arbitrary Hamiltonian as long as the Hamiltonian can be decomposed into a sum of local terms,
\begin{equation}
    H=h_1+...+h_N
\end{equation}
The first-order Suzuki-Trotter formula is given by
\begin{equation}
    e^{-iHt}=(e^{-ih_1\Delta t}\cdot ...\cdot e^{-ih_N\Delta t})^{t/\Delta t}+\mathcal{O}(t\Delta t)
\end{equation}
with an error depending on the step parameter $\Delta t$. Higher orders of the Suzuki-Trotter formula can be generalized through recursion, where the errors become smaller with each iteration, \cite{hatano2005suzuki}. The first-order Suzuki-Trotter approximation can be used to simulate, for instance, the exponential of the Hamiltonian $H=XX+YY+ZZ$,
\begin{equation}
\begin{split}
    e^{-iHt} \simeq (e^{-i\Delta XXt}e^{-i\Delta YYt}e^{-i\Delta ZZt})^{t/\Delta t}
\end{split}
\end{equation}
Through a quantum circuit involving a mix between the standard and the inverted staircase algorithms, it is possible to simulate such exponential, see figure \ref{fig:XXYYZZ}. 

\begin{figure*}[h]
        \centering
        \begin{tikzpicture}
		\draw (4.5,5.5)--(14.35,5.5);
            \draw (4.5,4.25)--(14.35,4.25);
            \filldraw[draw=black, fill=white] (5.25,4.5)--(6.55,4.5)--(6.55,4)--(5.25,4)--(5.25,4.5);
            \node at (5.9,4.25){$R_x\left(2\Delta t\right)$};

            \draw (4.85,5.5) circle (5pt);
            \filldraw (4.85,4.25) circle (2pt);
            \draw (4.85,5.65)--(4.85,4.25);

            \draw (7,5.5) circle (5pt);
            \filldraw (7,4.25) circle (2pt);
            \draw (7,5.65)--(7,4.25);

            \filldraw[draw=black, fill=white] (7.4,4.5)--(8,4.5)--(8,4)--(7.4,4)--(7.4,4.5);
            \node at (7.7,4.25){\large $\frac{\pi}{2}$};

            \filldraw[draw=black, fill=white] (7.4,5.75)--(8,5.75)--(8,5.25)--(7.4,5.25)--(7.4,5.75);
            \node at (7.7,5.5){\large $\frac{\pi}{2}$};

            \draw (8.4,5.5) circle (5pt);
            \filldraw (8.4,4.25) circle (2pt);
            \draw (8.4,5.65)--(8.4,4.25);

            \filldraw[draw=black, fill=white] (8.8,4.5)--(10.1,4.5)--(10.1,4)--(8.8,4)--(8.8,4.5);
            \node at (9.45,4.25){$R_x\left(2\Delta t\right)$};

            \draw (10.5,5.5) circle (5pt);
            \filldraw (10.5,4.25) circle (2pt);
            \draw (10.5,5.65)--(10.5,4.25);

            \filldraw[draw=black, fill=white] (10.9,4.5)--(11.5,4.5)--(11.5,4)--(10.9,4)--(10.9,4.5);
            \node at (11.2,4.25){\large $-\frac{\pi}{2}$};

            \filldraw[draw=black, fill=white] (10.9,5.75)--(11.5,5.75)--(11.5,5.25)--(10.9,5.25)--(10.9,5.75);
            \node at (11.2,5.5){\large $-\frac{\pi}{2}$};

            \draw (11.9,4.25) circle (5pt);
            \filldraw (11.9,5.5) circle (2pt);
            \draw (11.9,5.5)--(11.9,4.1);

            \filldraw[draw=black, fill=white] (12.3,4.5)--(13.6,4.5)--(13.6,4)--(12.3,4)--(12.3,4.5);
            \node at (12.95,4.25){$R_z\left(2\Delta t\right)$};

            \draw (14,4.25) circle (5pt);
            \filldraw (14,5.5) circle (2pt);
            \draw (14,5.5)--(14,4.1);

\end{tikzpicture}
	\caption{Quantum circuit simulating the exponential $e^{-iHt}$ of the Hamiltonian $H=XX+YY+ZZ$ up to certain error using the first-order Suzuki-Trotter formula. This decomposition involves a mix between the inverted staircase algorithm and the standard one amounting to six $\operatorname{CNOT}$ gates and seven one-qubit gates.}
	\label{fig:XXYYZZ}
\end{figure*}

Such Hamiltonians have direct applications in physics. In \cite{raeisi2012quantum}, similar quantum circuits are generated to simulate circuits that can construct the ground state of the Hamiltonian. Moreover, in \cite{yordanov2020efficient} a quantum circuit algorithm that performs single and double qubit excitations is constructed. Such excitations are created by typical creation and annihilation operators, which are translated into the language of quantum computation through the Jordan-Wigner transformations. 
All of the exponentials of any Hamiltonian can be constructed through the staircase algorithms and the Suzuki-Trotter formulas, however, for some examples, a specifically tailored algorithm, as is the case for the  circuit simulating single and double qubit excitations, might be more efficient, \cite{yordanov2020efficient}.

\section{Measurement}
\label{sec:measurement}
It is widely known that the phenomena occurring in quantum physics have a probabilistic nature. To that end, measurements are used in quantum circuits to obtain a tangible result by replacing quantum information with classical information. However, some of the information contained in the quantum state is "lost" after measurement, e.g. two different quantum circuits can produce the same outcomes after a particular measurement. 

Fermionic gates of the type we have defined give the same probability outcome of measurement as the non-fermionic quantum gates. To see this, let $\ket{\psi}$ be a general vector state on $SU(4)$:
\begin{equation} \ket{\psi}=\alpha_{00}\ket{00}+\alpha_{01}\ket{01}+\alpha_{10}\ket{10}+\alpha_{11}\ket{11}
\end{equation}
where we use the same basis as in section \ref{sec:gates}. Then, the action of a fermionic $\operatorname{SWAP}$ gate on $\ket{\psi}$ is given by
\begin{equation}
        \operatorname{fSWAP}\ket{\psi}=\alpha_{00}\ket{00}+\alpha_{10}\ket{01}+\alpha_{01}\ket{10}-\alpha_{11}\ket{11}
\end{equation}
where the relevant distinction with the non-fermionic $\operatorname{SWAP}$ gate lies on the last minus sign. 
It follows that after measuring one of the qubits, the probabilities of obtaining that measurement are the same as we had measured the state $\ket{\psi}$ after being acted with the standard $\operatorname{SWAP}$ gate instead. For instance, the probability that the first qubit is 0 is given by
\begin{equation}
    \begin{split}       |\bra{0}\operatorname{fSWAP}\ket{\psi}|^2=|\alpha_{00}|^2+|\alpha_{10}|^2=|\bra{0}\operatorname{SWAP}\ket{\psi}|^2
    \end{split}
\end{equation}
and, similarly, the probability that the second qubit is 1 is
\begin{equation}
    \begin{split} |\bra{1}\operatorname{fSWAP}\ket{\psi}|^2=|\alpha_{10}|^2+|\alpha_{11}|^2=|\bra{1}\operatorname{SWAP}\ket{\psi}|^2
    \end{split}
\end{equation}
giving thus the same probabilities as for the $\operatorname{SWAP}$ case. This is due to the absolute value of the probability amplitudes $\alpha_{ij}$ for the given state vector.  This absolute value is a direct consequence of Born's rule for quantum measurements \cite{Peres_2004}, i.e. probability outcomes of measurements are not affected by phases occurring in probability amplitudes, only by their absolute value. However, phases play an important role when interference effects are considered. For instance, Grover's algorithm, which is one of the most notable algorithms of quantum computing, relies on interference effects, \cite{grover1996fast}.

To illustrate the importance of interference effects for quantum measurements, let us discuss in detail the more general case for interference effects caused by the phase difference introduced by the fermionic $\operatorname{SWAP}$ for a 2-qubit system.

Similar as before, let $\ket{\psi}$ be an arbitrary 2-qubit state described as:

\begin{equation}
\ket{\psi}=\alpha_{00}\ket{00}+\alpha_{01}\ket{01}
        +\alpha_{10}\ket{10}+\alpha_{11}\ket{11}
\end{equation}

Then, the action of the $\operatorname{SWAP}$ and fermionic $\operatorname{SWAP}$ gates on this state is given respectively by the following:

\begin{equation}\label{eq:swapstate}
\operatorname{SWAP}\ket{\psi}=\alpha_{00}\ket{00}+\alpha_{10}\ket{01}
          +\alpha_{01}\ket{10}+\alpha_{11}\ket{11}
\end{equation}
And 

\begin{equation}\label{eq:fwapstate}  
\begin{split}
    \operatorname{fSWAP}\ket{\psi}=\alpha_{00}\ket{00}+\alpha_{10}\ket{01}   +\alpha_{01}\ket{10}-\alpha_{11}\ket{11}
\end{split}
\end{equation}
Since the Hadamard gate is a phase-shifting gate, we can use the composite Hadamard gate $H_{all} \equiv H \otimes H$ to study interference effects. The matrix representation of the composite Hadamard gate is given in the standard basis by:

\begin{equation}
    \text{H}_{all}:=H\otimes H\equiv \begin{pmatrix}
    1&1&1&1\\
    1&-1&1&-1\\
    1&1&-1&-1\\
    1&-1&-1&1\end{pmatrix}.
\end{equation}
We can now compute the action of this gate on the states given in (\ref{eq:swapstate}) and (\ref{eq:fwapstate}):
\begin{equation}\label{eq:swap}
\begin{split}
       H_{all}\left(\operatorname{SWAP}  \ket{\psi} \right)= \left( \frac{\alpha_{00}+\alpha_{01}+\alpha_{10}+\alpha_{11}}{2}\right) \ket{00} \\
       +\left( \frac{\alpha_{00}+\alpha_{01}
       -\alpha_{10}-\alpha_{11}}{2}\right)\ket{01} \\
       +\left( \frac{\alpha_{00}-\alpha_{01}+\alpha_{10}-\alpha_{11}}{2}\right) \ket{10} \\
       +\left( \frac{\alpha_{00}-\alpha_{01}-\alpha_{10}+\alpha_{11}}{2}\right) \ket{11} 
\end{split}
\end{equation}

\begin{equation}\label{eq:fswap}
\begin{split}
       H_{all} \left( \operatorname{fSWAP} \ket{\psi} \right)= \left( \frac{\alpha_{00}+\alpha_{01}+\alpha_{10}-\alpha_{11}}{2}\right) \ket{00}\\
       +\left( \frac{\alpha_{00}+\alpha_{01}-\alpha_{10}+\alpha_{11}}{2}\right)\ket{01} \\
       +\left( \frac{\alpha_{00}-\alpha_{01}+\alpha_{10}+\alpha_{11}}{2}\right) \ket{10}\\
       +\left( \frac{\alpha_{00}-\alpha_{01}-\alpha_{10}-\alpha_{11}}{2}\right) \ket{11} 
\end{split}
\end{equation}
Note that the coefficient $\alpha_{11}$ enters with an opposite sign in all the probability amplitudes of the final state, thus, altering the measurement outcome of both states.
This is just one instance where quantum interference leads to measurable effects in a quantum circuit, illustrating the importance of phase differences, such as the one introduced by fermionic gates, and should be generalizable to a $n$-qubit state $\ket{\psi}$ in a straightforward manner.

Let us now illustrate the interference behavior in a concrete example. Let $\ket{\phi}$ be a 2-qubit state given by :
\begin{equation}
    \ket{\phi}= \frac{1}{\sqrt{2}}\left ( \ket{00}+ \ket{11}\right).
\end{equation}

 Then, we can describe the action of the $\operatorname{SWAP}$ and $\operatorname{fSWAP}$ gates in this state as follows:
 
\begin{equation}
\begin{split}
        \operatorname{SWAP}\ket{\phi}=\frac{1}{\sqrt{2}}\left ( \ket{00}+ \ket{11}\right),
\end{split}
\end{equation}
and 

\begin{equation}
\begin{split}
        \operatorname{fSWAP}\ket{\phi}=\frac{1}{\sqrt{2}}\left ( \ket{00}-\ket{11}\right).
\end{split}
\end{equation}

We can compute the outcome of acting with the operator  $H_{all}  \equiv H \otimes  H$ on the left of the states above using 
\ref{eq:swap} and \ref{eq:fswap}, in the case where $\alpha_{00}= \alpha_{11}= \frac{1}{\sqrt{2}}$ and $\alpha_{10}= \alpha_{01}= 0$, giving:

\begin{equation} \label{eq:swapinterference}
     H_{all} \left(\operatorname{SWAP} \ket{\phi} \right)=\frac{1}{\sqrt{2}}\left ( \ket{00}+ \ket{11}\right),
\end{equation}
and 
\begin{equation}\label{eq:fswapinterference}
         H_{all} \left( \operatorname{fSWAP} \ket{\phi} \right) =\frac{1}{\sqrt{2}}\left ( \ket{01}+ \ket{10}\right).
\end{equation}

As we can see, the Haldamard gate causes the initial amplitudes $\alpha_{ij}$ of the state $\ket{\phi}$ to interfere, resulting in very different out states.

This demonstrates that in general one cannot simply substitute all $\operatorname{SWAP}$ gates on a quantum circuit with $\operatorname{fSWAPs}$ without altering the outcome of the circuit, since interference effects need to be considered. In particular, the modified staircase algorithm discussed here utilizes multiple phase-shifting gates, as was discussed in \ref{sec:inversestaircase}, therefore, interference effects will play a role in the outcomes and one must be careful to use the correct swapping gate for the problem. Similar reasoning works for the enlarged fermionic quantum gates.



\section{Discussion}
\label{sec:discussion}
The staircase algorithms are a straightforward approach to determining the quantum circuits simulating the exponential of an arbitrary Hamiltonian. As such, these have broad applications in simulating quantum physical systems. The inverted staircase algorithm is a polynomial improvement of the standard staircase algorithm in the number of one-qubit gates so that a hybrid algorithm between both the standard and the inverted staircase algorithms provides a significant improvement.

Fermionic $\operatorname{SWAP}$ gates are introduced to account for the minus signs that arise when two fermionic modes are exchanged. Although they were originally introduced to simulate strongly correlated quantum many-body systems, we showed that these gates could be systematically enlarged extending their use to a broader scope of algorithms. In this paper, we introduced fermionic $\operatorname{SWAP}$ gates to optimize the inverted staircase algorithm. These gates, if implementable, allow a polynomial improvement in the number of $\operatorname{CNOT}$ gates not only with respect to the staircase algorithms but also potentially with a variety of quantum circuits. 

Moreover, in the last section, we discussed the difference between non-fermionic and fermionic quantum gates. We showed that, due to Born's rule for quantum measurements, phases do not always play a role so one might substitute in certain cases fermionic gates for non-fermionic ones. We analyzed further the case in which interference phenomena are considered, where we showed a relevant distinction between both types of gates and thus the necessity of considering fermionic quantum gates.

The simulations of the exponentials of different Hamiltonians have broader applications in physics. However, most of the terms composing an arbitrary Hamiltonian do not commute so an approximation is necessary to simulate its exponential. Through the Suzuki-Trotter decomposition, we showed a straightforward method to simulate the exponential of any Hamiltonian up to an error. We used a hybrid staircase algorithm to simulate the dynamics of a standard quantum Hamiltonian and compute a quantum circuit. However, we noted that for specific examples such an approach might be too broad and not as efficient as desired.

In the last section, we have examined the difference between the probability outcomes of fermionic and non-fermionic quantum gates, see section \ref{sec:measurement}. Through considering a phenomenon such as interference, we have shown that one cannot, in general, substitute non-fermionic gates on a quantum circuit with fermionic gates without altering the outcome of the circuit. This implies that fermionic quantum gates are a necessary 

\section*{Acknowledgments}

The authors acknowledge funding by the German Federal Ministry for Research and Education (BMBF) under grant 13N16089 (BAIQO) of the funding program "QUantumtechnologien – von den Grundlagen zum Markt" (quantum technologies – from basic research to market).

\bibliographystyle{ieeetr}
\bibliography{localbibliography}

\end{document}